\documentclass[12pt]{article}
\usepackage{amsmath}
\usepackage{graphicx}
\usepackage{natbib}
\usepackage{url}

\def\spacingset#1{\renewcommand{\baselinestretch}
	{#1}\small\normalsize} \spacingset{1}

\newcommand{\iid}{\overset{\scriptscriptstyle\textrm{iid}}{\sim}}
\newcommand{\bm}[1]{\boldsymbol{#1}}
\newcommand{\trans}{\scriptscriptstyle \textrm{T}}
\newcommand{\ord}[1]{\scriptscriptstyle (#1)}
\newcommand{\iter}[1]{\scriptscriptstyle [#1]}
\newcommand{\sub}[1]{\scriptscriptstyle #1}

\title{\bf General P-Splines\\ for Non-Uniform B-Splines}
\author{
	Zheyuan Li\\
	School of Mathematics and Statistics, Henan University\\
	and\\
	Jiguo Cao\\
	Department of Statistics and Actuarial Science, Simon Fraser University}
\date{}

\begin{document}

\maketitle

\bigskip
\begin{abstract}
We proposed a new penalized B-splines estimator, the general P-spline, to accommodate non-uniform B-splines on unevenly spaced knots. It is a complement to Eilers and Marx's standard P-spline tailored for uniform B-splines on equidistant knots. At its core, we derived a novel general difference penalty that accounts for irregular knot spacing, while still being easy to compute and interpret. Both P-spline variants are useful for practical smoothing, because either one can produce a more satisfactory fit than the other, depending on the knot sequence being used and the data being analyzed. Therefore, practitioners should try out both before betting on either one, for which we have implemented general P-spline in \textbf{R} packages \textbf{gps} and \textbf{gps.mgcv}. The new general P-spline is closely related to O'Sullivan spline (or O-spline) through a sandwich formula that links general difference penalty to derivative penalty. Though both penalties seem equally powerful in wiggliness control for their mathematical association and statistical similarity, simulation studies show that general P-spline either outperforms O-spline in terms of mean squared error, or performs equally well, making it a superior replacement of O-spline.
\end{abstract}

\vfill

\newpage
\spacingset{1.5}

\section{Introduction}

P-spline \citep{P-splines,P-splines-20-years,P-splines-book} is a popular penalized B-splines estimator for univariate smooth functions. It has been applied in many statistical modeling frameworks, like generalized additive models \citep{WOS:000457464800009}, single-index models \citep{WOS:000440611000006}, generalized partially linear single-index models \citep{WOS:000395004300017}, functional mixed-effects models \citep{WOS:000444443000004}, survival models \citep{WOS:000641387600001,WOS:000364259800022}, models for longitudinal data \citep{WOS:000418746100004,WOS:000436403600033}, additive quantile regression models \citep{WOS:000549910800001}, varying coefficient models \citep{WOS:000450660500010}, quantile varying coefficient models \citep{WOS:000434068100003}, spatial models \citep{WOS:000448217800001,WOS:000426321800004}, spatiotemporal models \citep{WOS:000489511900001,WOS:000456529100005} and spatiotemporal quantile/expectile regression models \citep{WOS:000461592800005,WOS:000546038500003}.

\begin{figure}
	\centering
	\includegraphics[width = 0.7\columnwidth]{"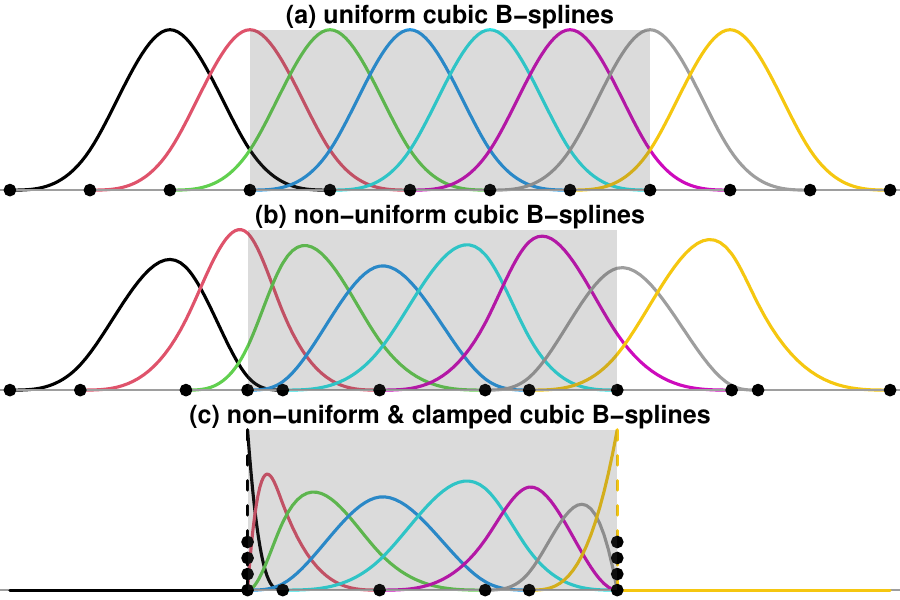"}
	\caption{Cubic B-splines classified by their knots (black dots, stacked if clamped): (a) uniform B-splines on equidistant knots; (b) non-uniform B-splines on unevenly spaced knots; (c) a special case of non-uniform B-splines with clamped boundary knots.}
	\label{fig1}
\end{figure}

In a nutshell, P-spline represents an unknown univariate function as a linear combination of B-splines \citep{deBoor-book}, whose coefficients are subject to a difference penalty to prevent overfitting. Although B-splines can be constructed on arbitrary knot sequences, P-spline is bundled with uniform B-splines on equidistant knots (see Figure \ref{fig1}(a)). As a result, to handle non-uniform B-splines on unevenly spaced knots (see Figure \ref{fig1}(b) and (c)), another type of penalized B-splines called O-spline \citep{O-splines-1}, characterized by a derivative penalty, is hitherto the only choice.

\begin{figure}
	\centering
	\includegraphics[width = 0.8\columnwidth]{"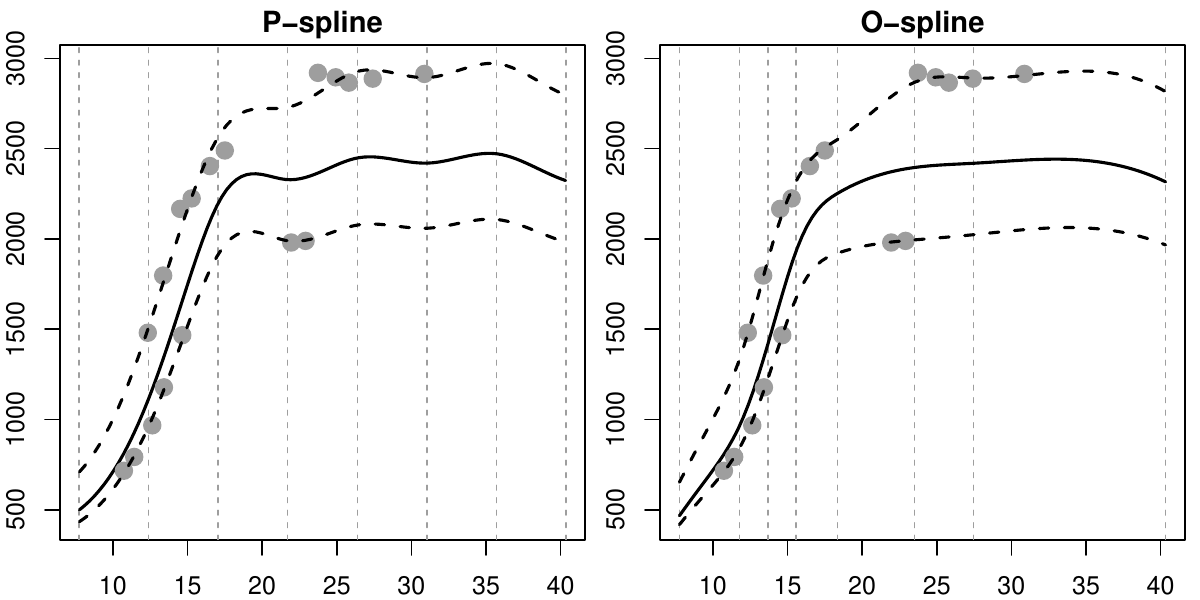"}
	\caption{Estimated population BMC trajectory (solid line) and subject BMC trajectories (dashed lines) for two selected subjects (gray dots). Left: cubic P-spline estimates with 10 uniform B-splines on equidistant knots (vertical dashed lines). Right: cubic O-spline estimates with 10 non-uniform B-splines on unevenly spaced knots (vertical dashed lines).}
	\label{fig2}
\end{figure}

Unevenly spaced knots can be more satisfactory than equidistant knots. For example, \cite{spectral-density-estimation-using-P-splines-with-quantile-knots} demonstrated that when estimating spectral densities with spikes, O-spline fit on a few unevenly spaced knots beats P-spline fit on a large number of equidistant knots, in terms of both goodness of it and computation time. As the motivating example of this paper, Figure \ref{fig2} shows that when smoothing annually measured bone mineral content (BMC, in grams) from 112 subjects, collected in the Pediatric Bone Mineral Accrual Study \citep{PBMAS-study-original}, trajectories estimated by P-splines on equidistant knots are suspiciously wiggly, whereas trajectories estimated by O-splines on unevenly spaced knots are plausibly smooth (see Section \ref{subsection: BMC longitudinal data} for details of BMC data modeling).

The primary aim of this paper is to generalize P-spline so that it can accommodate non-uniform B-splines. The newly developed P-spline will be termed the general P-spline. By contrast, the original P-spline will hereafter be referred to as the standard P-spline. See Table \ref{table1} for a summary of the capability of all relevant penalized B-splines estimators.

\begin{table}
	\caption{Capability of several penalized B-splines estimators}
	\label{table1}
	\centering
	\bigskip
	\begin{tabular}{|c|c|c|}
		\cline{2-3}
		\multicolumn{1}{c|}{} & derivative penalty & difference penalty\\
		\hline
		\parbox[t][\totalheight][s]{18.7ex}{uniform B-splines\\ (equidistant knots)} & O-spline (os) & \parbox[t][\totalheight][s]{22.6ex}{standard P-spline (sps)\\general P-spline (gps)}\\
		\hline
		\parbox[t][\totalheight][s]{23.6ex}{non-uniform B-splines\\(unevenly spaced knots)} & O-spline (os) & \parbox[t][\totalheight][s]{22.6ex}{general P-spline (gps)}\\
		\hline
	\end{tabular}
\end{table}

This paper has three major contributions. Firstly, we proposed and justified general P-spline, which is not a trivial task. Although it has been conjectured \citep{P-splines,Wood-GAMs-book,review-of-R-packages-on-splines-2019,spectral-density-estimation-using-P-splines-with-quantile-knots} that difference penalty should be calculated using weighed differences, possibly divided differences, when handling non-uniform B-splines, it is not intuitive at all how to do this because the number of B-spline coefficients does not equal the number of knots. We are the first to formulate the procedure, which is, in fact, not doing divided differences. Secondly, we derived a sandwich formula linking difference penalty to derivative penalty, based on which we explored the relation between the two penalties. This is enlightening, because the understanding of why difference penalty works for uniform B-splines, why it fails for non-uniform B-splines and how it compares with derivative penalty is very lacking in the literature, and the success of difference penalty is so far more a practical finding than a theoretical promise. We fill this gap, by establishing a one-to-one mathematical correspondence between the two penalties and showing their similarity in the Bayesian view (with evidence from simulations). In this way, the mechanism of difference penalty is no more magic. We also demonstrated through simulation studies that general P-spline either outperforms O-spline in terms of mean squared error (MSE), or performs equally well, making it a superior replacement of O-spline. Last but not least, we brought more insight into the debate \citep{SemiPar-article-2009, SemiPar-with-O-splines, P-splines-20-years} on the supremacy of either equidistant knots and unevenly spaced knots, by telling a two-sided story. On one hand, we showed that in well-designed simulation studies where several conditions are met, exact knot locations do not seem to matter, and general P-spline on unevenly spaced knots is competitive to standard P-spline on equidistant knots. On the other hand, in practical smoothing where any of the conditions could be violated, knot locations matter and either general P-spline or standard P-spline can produce a more satisfactory fit than the other. Therefore, it is generally a good idea for practitioners to try out both P-spline variants before betting on either one, for which we have implemented general P-spline via \textbf{R} packages \textbf{gps} \citep{gps-package} and \textbf{gps.mgcv} \citep{gps.mgcv-package}.

The rest of this paper is planned as follows. Section \ref{section: the new general P-spline} explains why standard P-spline fails for non-uniform B-splines, derives the new general P-spline and explores its connection with O-spline. Section \ref{section: simulation studies} conducts well-designed simulation studies to demonstrate that general P-spline either outperforms O-spline and standard P-spline, or performs equally well. Section \ref{section: real data examples} studies the impact of knot placement by smoothing BMC longitudinal data and fossil shell data, showing that the former favors general P-spline on unevenly spaced knots, while the latter favors standard P-spline on equidistant knots. Section Discussion summarizes key findings of this paper, showcases other use of the sandwich formula and discusses knot placement in a broader scope.

\section{The New General P-Spline}
\label{section: the new general P-spline}

A smoothing model for observations $(x_i, y_i)_{1}^{n}$ hypothesizes:
\begin{equation*}
	y_i = f(x_i) + e_i, \kern 2mm e_i \iid \textrm{N}(0, \sigma_e^2),
\end{equation*}
where $f(x)$ is a smooth function estimator and $e_i$ is an iid Gaussian error. A penalized B-splines estimator assumes $f(x)$ to be a spline under B-splines representation:
\begin{equation*}
	f(x) = \sum_{j = 1}^{p}B_j(x)\beta_j,
\end{equation*}
with coefficients $\bm{\beta} = (\beta_1, \beta_2, \ldots, \beta_p)^{\trans}$ for B-splines $(B_j(x))_1^p$ estimated by minimizing a penalized least squares objective:
\begin{equation*}
	\sum_{i = 1}^n\bigg[y_i - \sum_{j = 1}^{p}B_j(x_i)\beta_j\bigg]^2 + \lambda\cdot\textrm{PEN}(\bm{\beta}),
\end{equation*}
where the penalty function $\textrm{PEN}(\bm{\beta})$ is some wiggliness measure for $f(x)$. Several penalized B-splines estimators exist, like O-spline and standard/general P-splines listed in Table \ref{table1}, each of which is characterized by a form of $\textrm{PEN}(\bm{\beta})$. The non-negative smoothing parameter $\lambda$ trades off $f(x)$'s closeness to data for $f(x)$'s smoothness and plays a critical role. It is common practice to choose its optimal value that minimizes the leave-one-out generalized cross-validation (GCV) error \citep{Wahba-spline-models-book}, although other criteria like maximizing the restricted log-likelihood also exist \citep{Wood-GAMs-book}. Now to motivate general P-spline, let's first explain why standard P-spline fails for non-uniform B-splines.

\subsection{Standard P-Spline}

\begin{figure}
	\centering
	\includegraphics[width = \columnwidth]{"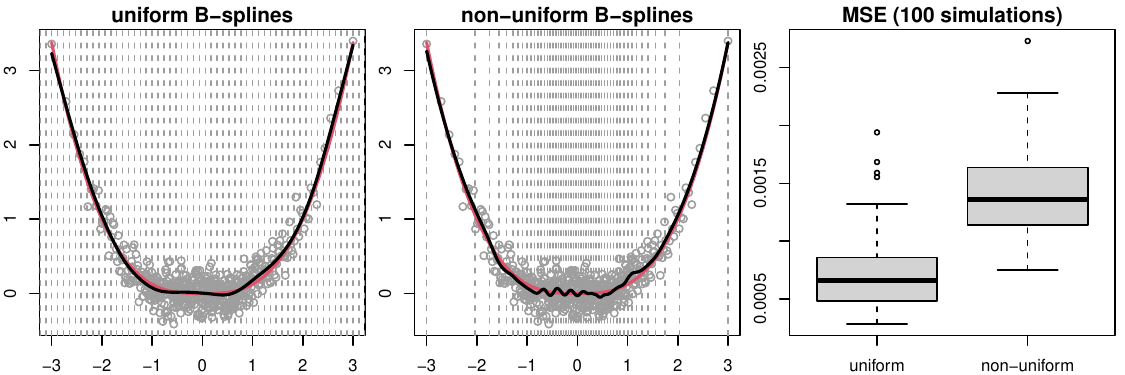"}
	\caption{Standard P-spline is tailored for uniform B-splines on equidistant knots, and does not make sense for non-uniform B-splines on unevenly spaced knots.}
	\label{fig3}
\end{figure}

For standard P-spline, the penalty function, hereafter called the standard difference penalty, is the sum of squared order-$m$ differences of nearby B-spline coefficients. To be precise, for $m = 1, 2, 3$, we have:
\begin{align*}
	\textrm{PEN}_{\textrm{sps}}^{\ord{1}}(\bm{\beta}) &= \sum_{j = 1}^{p - 1}(\beta_{j + 1} - \beta_j)^2,\\
	\textrm{PEN}_{\textrm{sps}}^{\ord{2}}(\bm{\beta}) &= \sum_{j = 1}^{p - 2}(\beta_{j + 2} - 2\beta_{j + 1} + \beta_j)^2,\\
	\textrm{PEN}_{\textrm{sps}}^{\ord{3}}(\bm{\beta}) &= \sum_{j = 1}^{p - 3}(\beta_{j + 3} - 3\beta_{j + 2} + 3\beta_{j + 1} - \beta_j)^3.
\end{align*}
Such penalty is tailored for uniform B-splines on equidistant knots, and does not make sense for non-uniform B-splines on unevenly spaced knots. To confirm this, consider fitting a standard cubic P-spline with a 2nd order penalty to 500 noisy observations of a U-shaped curve $y = \frac{1}{8}|x|^3$, where $x$-values are simulated from $\textrm{N}(0, 1)$ distribution. To construct uniform B-splines, we place 50 equidistant knots through the range of $x$-values. To construct non-uniform B-splines, we place 50 knots at equal quantiles of $x$-values. Figure \ref{fig3} shows that the fit under non-uniform B-splines representation is rather wiggly at the valley, while the fit under uniform B-splines representation is as smooth as it should be. The boxplot of MSE based on 100 simulations reassures that this phenomenon is persistent. In short, standard difference penalty fails to perform roughness control when handling non-uniform B-splines on unevenly spaced knots.

\begin{figure}[t]
	\centering
	\includegraphics[width = \columnwidth]{"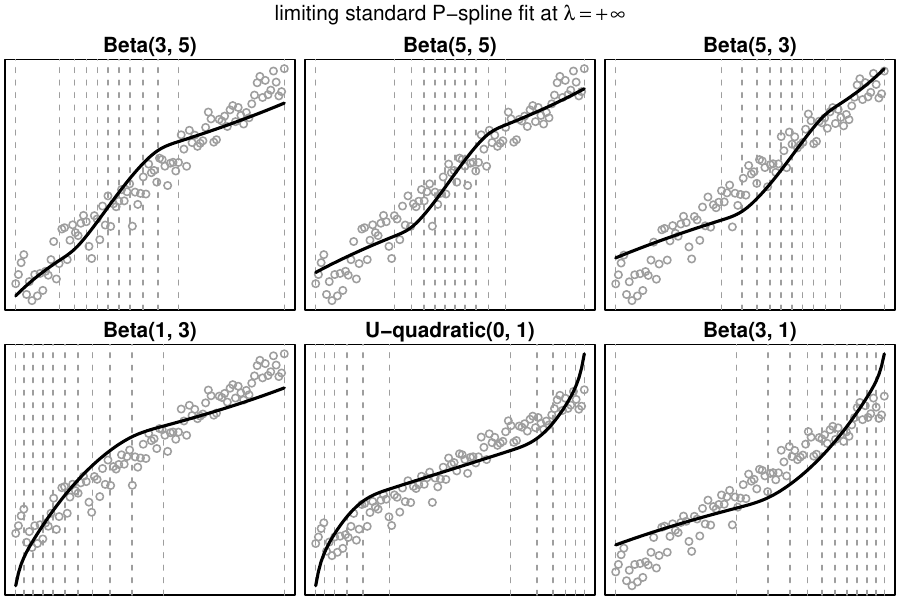"}
	\caption{Standard difference penalty is not a reliable wiggliness measure for spline function represented with non-uniform B-splines, because the limiting standard P-spline fit (solid line) at $\lambda = +\infty$ is inconsistent when knot locations (vertical dashed lines) vary.}		
	\label{fig4}
\end{figure}

There may be many explanations for such failure, but we focus on a particular flaw of standard difference penalty by investigating the limiting behavior of $f(x)$ at $\lambda = +\infty$. In this case, $\textrm{PEN}_{\textrm{sps}}^{\ord{m}}(\bm{\beta}) = 0$ and the P-spline fit lies in the penalty's null space. Taking $m = 2$ as an example, let's fit a standard P-spline set up with non-uniform cubic B-splines to noisy observations from $y = x$, $x \in [0, 1]$. To examine effects of knot locations, we attempt unevenly spaced knots that follow different distributions (primarily Beta distributions with varying shape parameters). Figure \ref{fig4} shows that the limiting fit at $\lambda = +\infty$ is sensitive to knot locations. Since unevenly spaced knots can be anywhere in general, this limiting fit is utterly unpredictable. Hence, the standard difference penalty is not a reliable roughness measure when handling non-uniform B-splines and will fail to control $f(x)$'s wiggliness as $\lambda$ varies on $(0, +\infty)$.

To overcome this defect, standard difference penalty must be upgraded to take uneven knot spacing into account. We call this generalization the general difference penalty, and it characterizes the new general P-spline. To make the construction of general difference penalty as transparent as possible, we present Section \ref{subsection: introducing general P-spline} as a tutorial by working through a small example and postpone its mathematical derivation to Section \ref{subsection: justifying general P-spline}. As we begin to work on knot sequences, we strongly advise readers to refer to the Appendix from time to time, which clarifies order ($d$), domain ($[a, b]$) and knots ($k$, $K$) of spline and B-splines.

\subsection{Introducing General P-Spline}
\label{subsection: introducing general P-spline}

Let
\begin{equation*}
	\bm{\Delta} =
	\begin{bmatrix}
		-1 & 1\\
		& -1 & 1\\
		& & \ddots & \ddots\\
		& & & -1 & 1
	\end{bmatrix}
\end{equation*}
be a fundamental difference matrix whose exact dimension can be inferred from the context, then the 1st, 2nd and 3rd order standard differences of $\bm{\beta}_0 = \bm{\beta}$ can be iteratively calculated as $\bm{\beta}_1 = \bm{\Delta}\bm{\beta}_0$, $\bm{\beta}_2 = \bm{\Delta}\bm{\beta}_1$ and $\bm{\beta}_3 = \bm{\Delta}\bm{\beta}_2$. Calculating general differences is no more complicated than an extra weighting process at each step. That is, instead of $\bm{\beta}_m = \bm{\Delta}\bm{\beta}_{m - 1}$, we compute $\bm{\beta}_m = \bm{W}_m^{-1}\bm{\Delta}\bm{\beta}_{m - 1}$ for some diagonal weight matrix $\bm{W}_m$. The key issue is just how to obtain $\bm{W}_m$ and we will make this crystal clear through the following example.

Consider non-uniform cubic B-splines ($d = 4$) on domain $[a, b] = [0, 4]$ with $k = 2$ interior knots $s_1 = 1$, $s_2 = 3$ and clamped boundary knots. In this case, there will be $p = k + d = 6$ B-splines and $K = p + d = 10$ knots. Coefficient vector $\bm{\beta}$ (or $\bm{\beta}_0$) has 6 elements and the complete knot sequence is:
\begin{center}
	\begin{tabular}{|c|c|c|c|c|c|c|c|c|c|}
		$t_1$ & $t_2$ & $t_3$ &$t_4$ &$t_5$ &$t_6$ &$t_7$ &$t_8$ & $t_9$ & $t_{10}$\\
		\hline
		0 & 0 & 0 & 0 & 1 & 3 & 4 & 4 & 4 & 4
	\end{tabular}
\end{center}
Now step by step, we show how to calculate $\bm{W}_1$, $\bm{W}_2$ and $\bm{W}_3$. Specifically, at each step, we identify a set of ``active'' knots and compute their differences at a correct lag.

(1) Lag-3 difference of the following ``active'' knots, divided by the lag, gives diagonal elements of $\bm{W}_1$:
\begin{center}
	\begin{tabular}{|c|c|c|c|c|c|c|c|c|c|}
		\rlap{$t_1$}{\textbackslash} & $t_2$ & $t_3$ &$t_4$ &$t_5$ &$t_6$ &$t_7$ &$t_8$ & $t_9$ & \rlap{$t_{10}$}{\textbackslash\textbackslash}\\
		\hline
		\rlap{0}{\textbackslash} & 0 & 0 & 0 & 1 & 3 & 4 & 4 & 4 & \rlap{4}{\textbackslash}
	\end{tabular}
\end{center}
\begin{equation*}
	\bm{W}_1 = \frac{1}{3}\begin{bmatrix}
		t_5 - t_2\\
		& t_6 - t_3\\
		& & t_7 - t_4\\
		& & & t_8 - t_5\\
		& & & & t_9 - t_6 
	\end{bmatrix} = \begin{bmatrix}
		\frac{1}{3}\\
		& \frac{3}{3}\\
		& & \frac{4}{3}\\
		& & & \frac{3}{3}\\
		& & & & \frac{1}{3}.
	\end{bmatrix}.
\end{equation*}

(2) Lag-2 difference of the following ``active'' knots, divided by the lag, gives diagonal elements of $\bm{W}_2$:
\begin{center}
	\begin{tabular}{|c|c|c|c|c|c|c|c|c|c|}
		\rlap{$t_1$}{\textbackslash} & \rlap{$t_2$}{\textbackslash} & $t_3$ & $t_4$ & $t_5$ & $t_6$ & $t_7$ & $t_8$ & \rlap{$t_9$}{\textbackslash} & \rlap{$t_{10}$}{\textbackslash\textbackslash}\\
		\hline
		\rlap{0}{\textbackslash} & \rlap{0}{\textbackslash} & 0 & 0 & 1 & 3 & 4 & 4 & \rlap{4}{\textbackslash} & \rlap{4}{\textbackslash}
	\end{tabular}
\end{center}
\begin{equation*}
	\bm{W}_2 = \frac{1}{2}\begin{bmatrix}
		t_5 - t_3\\
		& t_6 - t_4\\
		& & t_7 - t_5\\
		& & & t_8 - t_6
	\end{bmatrix} = \begin{bmatrix}
		\frac{1}{2}\\
		& \frac{3}{2}\\
		& & \frac{3}{2}\\
		& & & \frac{1}{2}
	\end{bmatrix}.
\end{equation*}

(3) Lag-1 difference of the following ``active'' knots, divided by the lag, gives diagonal elements of $\bm{W}_3$:
\begin{center}
	\begin{tabular}{|c|c|c|c|c|c|c|c|c|c|}
		\rlap{$t_1$}{\textbackslash} & \rlap{$t_2$}{\textbackslash} & \rlap{$t_3$}{\textbackslash} & $t_4$ & $t_5$ & $t_6$ & $t_7$ & \rlap{$t_8$}{\textbackslash} & \rlap{$t_9$}{\textbackslash} & \rlap{$t_{10}$}{\textbackslash\textbackslash}\\
		\hline
		\rlap{0}{\textbackslash} & \rlap{0}{\textbackslash} & \rlap{0}{\textbackslash} & 0 & 1 & 3 & 4 & \rlap{4}{\textbackslash} & \rlap{4}{\textbackslash} & \rlap{4}{\textbackslash}
	\end{tabular}
\end{center}
\begin{equation*}
	\bm{W}_2 = \frac{1}{1}\begin{bmatrix}
		t_5 - t_4\\
		& t_6 - t_5\\
		& & t_7 - t_6
	\end{bmatrix} = \begin{bmatrix}
		1\\ & 2\\ & & 1
	\end{bmatrix}.
\end{equation*}

We now reach the end of the iteration because there is no lag-0 difference for calculating $\bm{W}_4$. As a result, $\bm{\beta}_1 = \bm{W}_1^{-1}\bm{\Delta}\bm{\beta}_0$, $\bm{\beta}_2 = \bm{W}_2^{-1}\bm{\Delta}\bm{\beta}_1$ and $\bm{\beta}_3 = \bm{W}_3^{-1}\bm{\Delta}\bm{\beta}_2$. By expanding the recursion:
\begin{equation*}
	\begin{split}
		\bm{\beta}_1 &= \overbrace{\bm{W}_1^{-1}\bm{\Delta}}^{\bm{D}_1}\bm{\beta}_0,\\
		\bm{\beta}_2 &= \bm{W}_2^{-1}\bm{\Delta}\bm{\beta}_1 = \overbrace{\bm{W}_2^{-1}\bm{\Delta}\bm{W}_1^{-1}\bm{\Delta}}^{\bm{D}_2}\bm{\beta}_0,\\
		\bm{\beta}_3 &= \bm{W}_3^{-1}\bm{\Delta}\bm{\beta}_2 = \overbrace{\bm{W}_3^{-1}\bm{\Delta}\bm{W}_2^{-1}\bm{\Delta}\bm{W}_1^{-1}\bm{\Delta}}^{\bm{D}_3}\bm{\beta}_0,
	\end{split}
\end{equation*}
we can work out the 1st, 2nd and 3rd order general difference matrices:
\begin{gather*}
	\bm{D}_1 = \begin{bmatrix}
		-3 & 3\\
		& -1 & 1\\
		& & -\frac{3}{4} & \frac{3}{4}\\
		& & & -1 & 1\\
		& & & & -3 & 3
	\end{bmatrix},\\
	\bm{D}_2 = \begin{bmatrix}
		6 & -8 & 2\\
		& \frac{2}{3} & -\frac{7}{6} & \frac{1}{2}\\
		& & \frac{1}{2} & -\frac{7}{6} & \frac{2}{3}\\
		& & & 2 & -8 & 6
	\end{bmatrix},\kern 2mm
	\bm{D}_3 = \begin{bmatrix}
		-6 & \frac{26}{3} & -\frac{19}{6} & \frac{1}{2}\\
		& -\frac{1}{3} & \frac{5}{6} & -\frac{5}{6} & \frac{1}{3}\\
		& & -\frac{1}{2} & \frac{19}{6} & -\frac{26}{3} & 6
	\end{bmatrix}.
\end{gather*}
Meanwhile, deliberately setting $\bm{W}_1$, $\bm{W}_2$ and $\bm{W}_3$ to identity matrices yields standard difference matrices of order 1 to 3:
\begin{gather*}
	\bm{\Delta}_1 = \begin{bmatrix}
		-1 & 1\\
		& -1 & 1\\
		& & -1 & 1\\
		& & & -1 & 1\\
		& & & & -1 & 1
	\end{bmatrix},\\
	\bm{\Delta}_2 = \begin{bmatrix}
		1 & -2 & 1\\
		& 1 & -2 & 1\\
		& & 1 & -2 & 1\\
		& & & 1 & -2 & 1
	\end{bmatrix},\kern 2mm
	\bm{\Delta}_3 = \begin{bmatrix}
		-1 & 3 & -3 & 1\\
		& -1 & 3 & -3 & 1\\
		& & -1 & 3 & -3 & 1
	\end{bmatrix}.
\end{gather*}

It is easy to verify that standard difference penalty can be concisely expressed as:
\begin{equation*}
	\textrm{PEN}_{\textrm{sps}}^{\ord{m}}(\bm{\beta}) = \|\bm{\Delta}_m\bm{\beta}\|^2.
\end{equation*}
Thus, a standard P-spline estimator is the minimizer of:
\begin{equation*}
	\sum_{i = 1}^n\bigg[y_i - \sum_{j = 1}^pB_j(x)\beta_j\bigg]^2 + \lambda\|\bm{\Delta}_m\bm{\beta}\|^2.
\end{equation*}
In an analogy, we define the following general difference penalty:
\begin{equation*}
	\textrm{PEN}_{\textrm{gps}}^{\ord{m}}(\bm{\beta}) = \|\bm{D}_m\bm{\beta}\|^2.
\end{equation*}
and consequently, a general P-spline estimator minimizes:
\begin{equation*}
	\sum_{i = 1}^n\bigg[y_i - \sum_{j = 1}^pB_j(x)\beta_j\bigg]^2 + \lambda\|\bm{D}_m\bm{\beta}\|^2.
\end{equation*}
The two objectives differ apparently only in the use of difference matrix, but standard P-spline is tailored for uniform B-splines on equidistant knots, whereas general P-spline is well defined for non-uniform B-splines on unevenly spaced knots. Notably, $\bm{D}_m$ and $\bm{\Delta}_m$ have the same band sparsity, but the latter has constant diagonals, while the former does not due to the weighting procedure. Had knots been equidistant, all weight matrices would be proportional to identity matrices, so that $\bm{D}_m$ would be proportional to $\bm{\Delta}_m$, establishing the equivalence between two estimators. In short, general P-spline comprises standard P-spline as its special case, and can handle B-splines on any knot sequences.

\subsection{Justifying General P-Spline}
\label{subsection: justifying general P-spline}

\begin{figure}
	\centering
	\includegraphics[width = 0.6\columnwidth]{"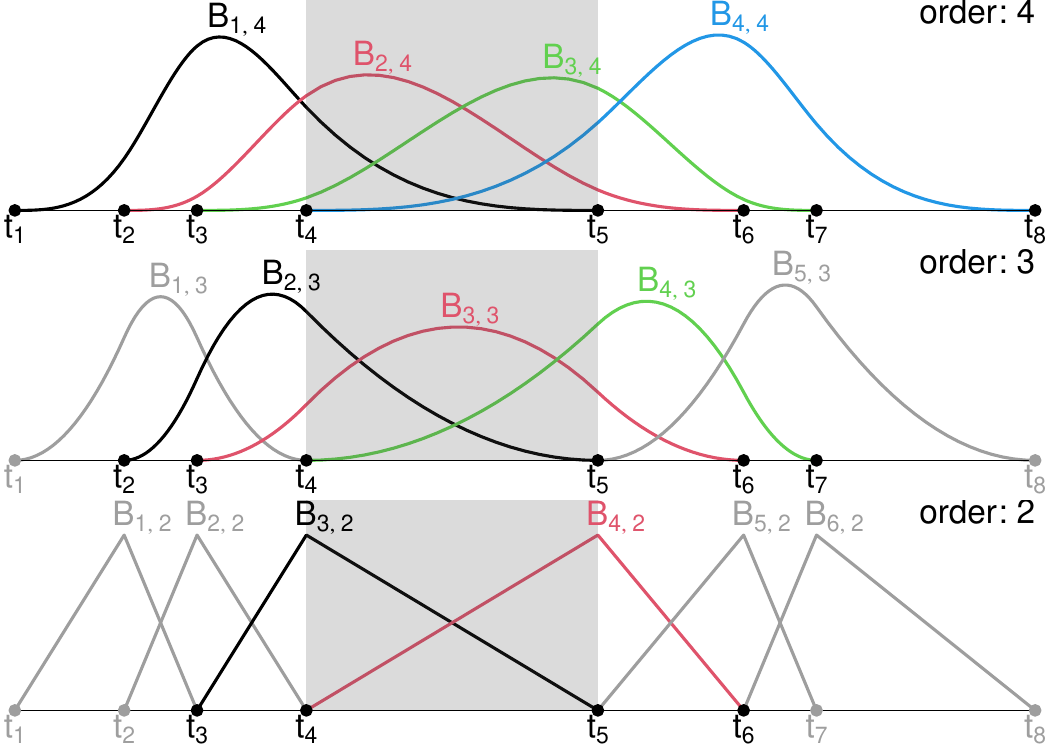"}
	\caption{On unevenly spaced knots $(t_j)_1^8$ (dots), constant B-splines $(B_{j,1}(x))_{1}^{7}$ (not drawn), linear B-splines $(B_{j,2}(x))_{1}^{6}$, quadratic B-splines $(B_{j,3}(x))_{1}^{5}$ and cubic B-splines $(B_{j,4}(x))_{1}^{4}$ can be sequentially constructed. The domain for a cubic spline is $[a, b] = [t_4, t_5]$ (shaded area). Low-order B-splines outside this domain are colored gray.}
	\label{fig5}
\end{figure}

We now justify the construction of general difference penalty exemplified in the previous section and interpret this new penalty. Mathematical derivation requires technical details on B-splines, and let's first explain their recursive construction. In general, on a full knot sequence $(t_j)_1^K$, we begin with $(K - 1)$ order-1 B-splines:
\begin{equation*}
	B_{j, 1}(x) =
	\begin{cases}
		1 & x \in [t_j, t_{j + 1}]\\
		0 & \textrm{otherwise},
	\end{cases}
\end{equation*}
and iteratively construct higher-order B-splines using:
\begin{equation*}
	B_{j, d}(x) = (x - t_j)\frac{B_{j, d - 1}(x)}{t_{j + d - 1} - t_j} + (t_{j + d} - x)\frac{B_{j + 1, d - 1}(x)}{t_{j + d} - t_{j + 1}}.
\end{equation*}
An illustration of such process is given in Figure \ref{fig5}, where from bottom to top, constant, linear, quadratic and cubic B-splines are sequentially constructed on unevenly spaced knots $(t_j)_1^8$. Clearly, B-splines have local support, that is, an order-$d$ B-spline is nonzero only on $d$ consecutive intervals.

Any order-$d$ spline $f(x)$ can be expressed as a linear combination of B-splines of the same order:
\begin{equation*}
	f(x) = \sum_{j = 1}^{p}B_{j,d}(x)\beta_{j,0},
\end{equation*}
so its 1st derivative is:
\begin{equation*}
	f'(x) = \sum_{j = 1}^{p}B_{j,d}'(x)\beta_{j,0}.
\end{equation*}
According to \cite{deBoor-book}, it holds that:
\begin{equation}
	\label{eqn: 1st derivative of B-spline}
	B_{j,d}'(x) = (d - 1)\left(\frac{B_{j, d - 1}(x)}{t_{j + d - 1} - t_j} - \frac{B_{j + 1, d - 1}(x)}{t_{j + d} - t_{j + 1}}\right).
\end{equation}
Therefore, the 1st derivative becomes:
\begin{equation*}
	\frac{f'(x)}{d - 1} = \sum_{j = 1}^{p}\frac{B_{j, d - 1}(x)}{t_{j + d - 1} - t_j}\beta_{j,0} - \sum_{j = 1}^{p}\frac{B_{j + 1, d - 1}(x)}{t_{j + d} - t_{j + 1}}\beta_{j,0}.
\end{equation*}
Here, the first term $B_{1, d - 1}(x)$ in the first summation and the last term $B_{p + 1, d - 1}(x)$ in the second summation are in fact zeros, because their supports are outside the domain (check Figure \ref{fig5} if this is not evident, watching for those low-order B-splines in gray). As a result,
\begin{equation*}
	\begin{split}
		\frac{f'(x)}{d - 1} &= \sum_{j = 2}^{p}\frac{B_{j, d - 1}(x)}{t_{j + d - 1} - t_j}\beta_{j,0} - \sum_{j = 1}^{p - 1}\frac{B_{j + 1, d - 1}(x)}{t_{j + d} - t_{j + 1}}\beta_{j,0}\\
		&= \sum_{j^{\star} = 1}^{p - 1}\frac{B_{j^{\star} + 1, d - 1}(x)}{t_{j^{\star} + d} - t_{j^{\star} + 1}}\beta_{j^{\star} + 1,0} - \sum_{j = 1}^{p - 1}\frac{B_{j + 1, d - 1}(x)}{t_{j + d} - t_{j + 1}}\beta_{j,0}\\
		&= \sum_{j = 1}^{p - 1}B_{j + 1, d - 1}(x)\frac{\beta_{j + 1,0} - \beta_{j,0}}{t_{j + d} - t_{j + 1}},
	\end{split}
\end{equation*}
where a change of variable $j = j^{\star} + 1$ is applied to the first summation from line 1 to 2. Alternatively, the result may be expressed as:
\begin{subequations}
	\begin{gather}
		\label{eqn: 1st derivative of f(x) -- a}
		\beta_{j + 1,1} = \frac{\beta_{j + 1,0} - \beta_{j,0}}{(t_{j + d} - t_{j + 1})/(d - 1)},\\
		\label{eqn: 1st derivative of f(x) -- b}
		f'(x) = \sum_{j = 1}^{p - 1}B_{j + 1, d - 1}(x)\beta_{j + 1,1}.
	\end{gather}
\end{subequations}
For 2nd derivative, we further differentiate \eqref{eqn: 1st derivative of f(x) -- b}:
\begin{equation*}
	f''(x) = \sum_{j = 1}^{p - 1}B_{j + 1, d - 1}'(x)\beta_{j + 1,1}.
\end{equation*}
Substituting $j$ with $j + 1$ and $d$ with $d - 1$ in \eqref{eqn: 1st derivative of B-spline} gives:
\begin{equation*}
	B_{j + 1, d - 1}'(x) = (d - 2)\left(\frac{B_{j + 1, d - 2}(x)}{t_{j + d - 1} - t_{j + 1}} - \frac{B_{j + 2, d - 2}(x)}{t_{j + d} - t_{j + 2}}\right).
\end{equation*}
In the same way as above, we can show:
\begin{equation*}
	\frac{f''(x)}{d - 2} = \sum_{j = 1}^{p - 2}B_{j + 2, d - 2}(x)\frac{\beta_{j + 2,1} - \beta_{j + 1,1}}{t_{j + d} - t_{j + 2}},
\end{equation*}
which can be rearranged as:
\begin{subequations}
	\begin{gather}
		\label{eqn: 2nd derivative of f(x) -- a}
		\beta_{j + 2, 2} = \frac{\beta_{j + 2,1} - \beta_{j + 1,1}}{(t_{j + d} - t_{j + 2})/(d - 2)},\\
		\label{eqn: 2nd derivative of f(x) -- b}
		f''(x) = \sum_{j = 1}^{p - 2}B_{j + 2, d - 2}(x)\beta_{j + 2, 2}.
	\end{gather}
\end{subequations}
Finally, applying induction to \eqref{eqn: 1st derivative of f(x) -- a}\eqref{eqn: 1st derivative of f(x) -- b}\eqref{eqn: 2nd derivative of f(x) -- a}\eqref{eqn: 2nd derivative of f(x) -- b}, we obtain the result for $f^{\ord{m}}(x)$ ($1 \leq m \leq d - 1$):
\begin{subequations}
	\begin{gather}
		\label{eqn: recursive coefficient difference - scalar form}
		\beta_{j + m, m} = \frac{\beta_{j + m, m - 1} - \beta_{j + m - 1, m - 1}}{(t_{j + d} - t_{j + m})/(d - m)},\\
		\label{eqn: m-th derivative of f(x) - scalar form}
		f^{\ord{m}}(x) = \sum_{j = 1}^{p - m}B_{j + m, d - m}(x)\beta_{j + m, m}.
	\end{gather}
\end{subequations}
To get rid of distracting subscripts in these equations, let's define:
\begin{gather*}
	\bm{B}_{d - m}(x) = \begin{bmatrix}
		B_{1 + m, d - m}(x) \\ B_{2 + m, d - m}(x) \\ \vdots \\ B_{p, d - m}(x)
	\end{bmatrix},\kern 2mm
	\bm{\beta}_m = \begin{bmatrix}
		\beta_{1 + m, m} \\ \beta_{2 + m, m} \\ \vdots \\ \beta_{p, m}
	\end{bmatrix},\\
	\bm{W}_m = \frac{1}{d - m}\begin{bmatrix}
		t_{d + 1} - t_{1 + m} \\ & t_{d + 2} - t_{2 + m} \\ & & \ddots \\ & & & t_{p + d - m} - t_{p}
	\end{bmatrix},
\end{gather*}
then \eqref{eqn: recursive coefficient difference - scalar form}\eqref{eqn: m-th derivative of f(x) - scalar form} can be written in a neat matrix-vector form:
\begin{subequations}
	\begin{align}
		\label{eqn: recursive coefficient difference - matrix form}
		\bm{\beta}_m &= \bm{W}_m^{-1}\bm{\Delta}\bm{\beta}_{m - 1},\\
		\label{eqn: m-th derivative of f(x) - matrix form}
		f^{\ord{m}}(x) &= \bm{B}_{d - m}(x)^{\trans}\bm{\beta}_m.
	\end{align}
\end{subequations}
By expanding the recursion in \eqref{eqn: recursive coefficient difference - matrix form}:
\begin{equation*}
	\begin{split}
		\bm{\beta}_m &= \bm{W}_m^{-1}\bm{\Delta}\bm{\beta}_{m - 1}\\
		& = \bm{W}_m^{-1}\bm{\Delta}\bm{W}_{m - 1}^{-1}\bm{\Delta}\bm{\beta}_{m - 2}\\
		& = \cdots = \underbrace{\bm{W}_m^{-1}\bm{\Delta}\bm{W}_{m - 1}^{-1}\bm{\Delta}\cdots\bm{W}_1^{-1}\bm{\Delta}}_{\bm{D}_m}\bm{\beta}_{0},
	\end{split}
\end{equation*}
we can work out the order-$m$ general difference matrix $\bm{D}_m$, so that:
\begin{align*}
	\bm{\beta}_m &= \bm{D}_m\bm{\beta}_0 = \bm{D}_m\bm{\beta},\\
	\textrm{PEN}_{\textrm{gps}}^{\ord{m}}(\bm{\beta}) &= \|\bm{D}_m\bm{\beta}\|^2 = \|\bm{\beta}_m\|^2.
\end{align*}

What we do in the previous section is exactly as same as above. Referring to Figure \ref{fig5}, we can interpret above results as follows. ``Active'' knots, or $(t_j)_{1 + m}^{K - m}$, are knots that are not colored gray in the Figure. The lag-$(d - m)$ difference of ``active'' knots, divided by $d - m$, gives diagonal elements of the weight matrix $\bm{W}_m$. Order-$(d - m)$ B-splines defined on ``active'' knots, i.e., B-splines that are not colored gray in the Figure, are collected in $\bm{B}_{d - m}(x)$. In particular, their linear combination with coefficients $\bm{\beta}_m$ gives $f^{\ord{m}}(x)$. Therefore, general difference penalty is the squared L\textsubscript{2} norm of $f^{\ord{m}}(x)$'s B-spline coefficients. When $\lambda = +\infty$, we have:
\begin{equation*}
	\textrm{PEN}_{\textrm{gps}}^{\ord{m}}(\bm{\beta}) = 0
	\kern 2mm \Leftrightarrow \kern 2mm
	\bm{\beta}_m = \bm{0}
	\kern 2mm \Leftrightarrow \kern 2mm
	f^{\ord{m}}(x) = 0,
\end{equation*}
which holds regardless knot locations. Thus, the limiting fit of general P-spline is always an order-$m$ polynomial.

\subsection{Connection with O-Spline}

Standard P-spline was proposed as an easier alternative to O-spline with derivative penalty:
\begin{equation*}
	\textrm{PEN}_{\textrm{os}}^{\ord{m}}(\bm{\beta}) = \int_{a}^{b} f^{\ord{m}}(x)^2\textrm{d}x = \bm{\beta^{\trans}}\bm{S}_m\bm{\beta},
\end{equation*}
where $\bm{S}_m$ is a symmetric matrix whose $(u, v)$\textsuperscript{th} element is $\int_{a}^{b}B_u^{\ord{m}}(x)B_v^{\ord{m}}(x)\textrm{d}x$. Back in the 1990s, it was increasingly difficult to compute $\bm{S}_m$ as $m$ grows, whereas it is always straightforward to set up standard difference matrix $\bm{\Delta}_m$. However, standard P-spline is not a full replacement of O-spline due to its incapability to handle non-uniform B-splines. Now with the advent of general P-spline, O-spline may be superseded in the near future, provided that the link between general difference penalty and derivative penalty is well understood. In this section, we compare two penalties from a few aspects to enhance our understanding of them.

Note that it is no longer challenging to compute $\bm{S}_m$. Using the algorithm of \cite{mgcv-B-splines}, it can be computed both exactly and efficiently. But to explore the connection between two penalties, we start with a new way to compute this matrix. Plugging $\bm{\beta}_m = \bm{D}_m\bm{\beta}$ into \eqref{eqn: m-th derivative of f(x) - matrix form}, we get:
\begin{equation*}
	f^{\ord{m}}(x) = \bm{B}_{d - m}(x)^{\trans}\bm{D}_m\bm{\beta}.
\end{equation*}
It follows that:
\begin{equation*}
	\textrm{PEN}_{\textrm{os}}^{\ord{m}}(\bm{\beta})
	= \int_{a}^{b}f^{\ord{m}}(x)^{\trans}f^{\ord{m}}(x)\textrm{d}x
	= \bm{\beta}^{\trans}\bm{D}_m^{\trans}\bm{\bar{S}}_m\bm{D}_m\bm{\beta},
\end{equation*}
where $\bm{\bar{S}}_m$ is a positive definite matrix whose $(u, v)$\textsuperscript{th} element is $\int_{a}^{b}B_{u, d - m}(x)B_{v, d - m}(x)\textrm{d}x$. Computation of $\bm{\bar{S}}_m$ can also employ Wood's algorithm, except that there is now no need to evaluate derivatives of B-splines. The general difference matrix $\bm{D}_m$ is also routine to compute (see Section \ref{subsection: introducing general P-spline}), thus, we may compute $\bm{S}_m$ from the sandwich formula:
\begin{equation*}
	\bm{S}_m = \bm{D}_m^{\trans}\bm{\bar{S}}_m\bm{D}_m.
\end{equation*}
The formula explicitly links two penalties together. Expressing general difference penalty as:
\begin{equation*}
	\textrm{PEN}_{\textrm{gps}}^{\ord{m}}(\bm{\beta}) = \|\bm{D}_m\bm{\beta}\|^2 = \bm{\beta^{\trans}}\bm{D}_m^{\trans}\bm{D}_m\bm{\beta},
\end{equation*}
we observe that it would equal derivative penalty if $\bm{\bar{S}}_m$ were an identity matrix. So how far is $\bm{\bar{S}}_m$ from being exactly or proportional to identity? Consider the example knot sequence used in Section \ref{subsection: introducing general P-spline} again. For cubic B-splines ($d = 4$), it can be computed that:
\begin{equation*}
	\bm{\bar{S}}_1 = \begin{bmatrix}
		\frac{1}{5} & \frac{11}{90} & \frac{2}{15}\\
		\frac{11}{90} & \frac{2}{5} & \frac{1}{5} & \frac{1}{135}\\
		\frac{2}{15} & \frac{1}{5} & \frac{22}{27} & \frac{1}{5} & \frac{2}{15}\\
		& \frac{1}{135} & \frac{1}{5} & \frac{2}{5} & \frac{11}{90}\\
		& & \frac{2}{15} & \frac{11}{90} & \frac{1}{5}
	\end{bmatrix},\kern 2mm \bm{\bar{S}}_2 = \begin{bmatrix}
		\frac{1}{3} & \frac{1}{6}\\
		\frac{1}{6} & 1 & \frac{1}{3}\\
		& \frac{1}{3} & 1 & \frac{1}{6}\\
		& & \frac{1}{6} & \frac{1}{3}
	\end{bmatrix},\kern 2mm
	\bm{\bar{S}}_3 = \begin{bmatrix}
		1\\ & 2\\ & & 1
	\end{bmatrix}.
\end{equation*}
In general, the bigger $m$ is, the closer $\bm{\bar{S}}_m$ is to diagonal. It is diagonal when $m = d - 1$, but not proportional to identity unless all knots are equidistant. Therefore, general difference penalty, taking $\bm{\bar{S}}_m$ to be identity, is almost always a simplification of derivative penalty.

Does the distinction between two penalties matter? Defining $\bm{K}_m = \bm{U}_m\bm{D}_m$, where $\bm{U}_m$ is $\bm{\bar{S}}_m$'s upper triangular Cholesky factor such that $\bm{\bar{S}}_m = \bm{U}_m^{\trans}\bm{U}_m$, we can express derivative penalty as:
\begin{equation*}
	\textrm{PEN}_{\textrm{os}}^{\ord{m}}(\bm{\beta}) = \|\bm{U}_m\bm{D}_m\bm{\beta}\|^2 = \|\bm{K}_m\bm{\beta}\|^2.
\end{equation*}
Because $\bm{U}_m$ has full rank, $\bm{D}_m$ and $\bm{K}_m$ have the same null space. Decomposing $\bm{\beta} = \bm{\xi} + \bm{\theta}$, where $\bm{\xi}$ is $\bm{\beta}$'s projection on this null space and $\bm{\theta}$ is orthogonal to $\bm{\xi}$, we have $\bm{D}_m\bm{\beta} = \bm{D}_m\bm{\theta}$, $\bm{K}_m\bm{\beta} = \bm{K}_m\bm{\theta}$ and $\bm{D}_m\bm{\xi} = \bm{K}_m\bm{\xi} = \bm{0}$. In other words, both penalties act on $\bm{\theta}$ only, so we may instead write:
\begin{align*}
	\textrm{PEN}_{\textrm{gps}}^{\ord{m}}(\bm{\theta}) &= \|\bm{D}_m\bm{\theta}\|^2,\\
	\textrm{PEN}_{\textrm{os}}^{\ord{m}}(\bm{\theta}) &= \|\bm{K}_m\bm{\theta}\|^2.
\end{align*}
Since $\bm{U}_m$ has full rank, for any $\bm{\theta}_{\textrm{gps}}$, there is a unique $\bm{\theta}_{\textrm{os}}$ such that $\bm{D}_m\bm{\theta}_{\textrm{gps}} = \bm{K}_m\bm{\theta}_{\textrm{os}}$ and $\textrm{PEN}_{\textrm{gps}}^{\ord{m}}(\bm{\theta}_{\textrm{gps}}) = \textrm{PEN}_{\textrm{os}}^{\ord{m}}(\bm{\theta}_{\textrm{os}})$, establishing a one-to-one correspondence between two functions. Therefore, while two penalties do not equal each other for a common $\bm{\theta}$, they are equivalent in wiggliness control. As $\lambda \to +\infty$, both O-spline and general P-spline tend to the same least squares order-$m$ polynomial. As $\lambda \to 0$, both estimators tend to the same least squares regression spline.

\begin{figure}
	\centering
	\includegraphics[width = 0.7\columnwidth]{"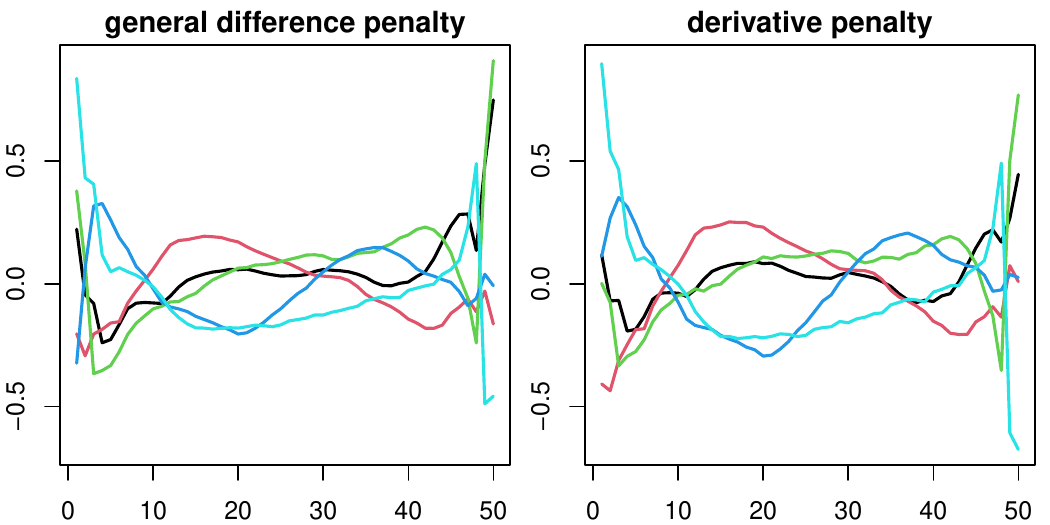"}
	\caption{Prior B-spline coefficients related to general difference penalty and derivative penalty are very similar.}
	\label{fig6}
\end{figure}

\begin{figure}[ht]
	\centering
	\includegraphics[width = 0.7\columnwidth]{"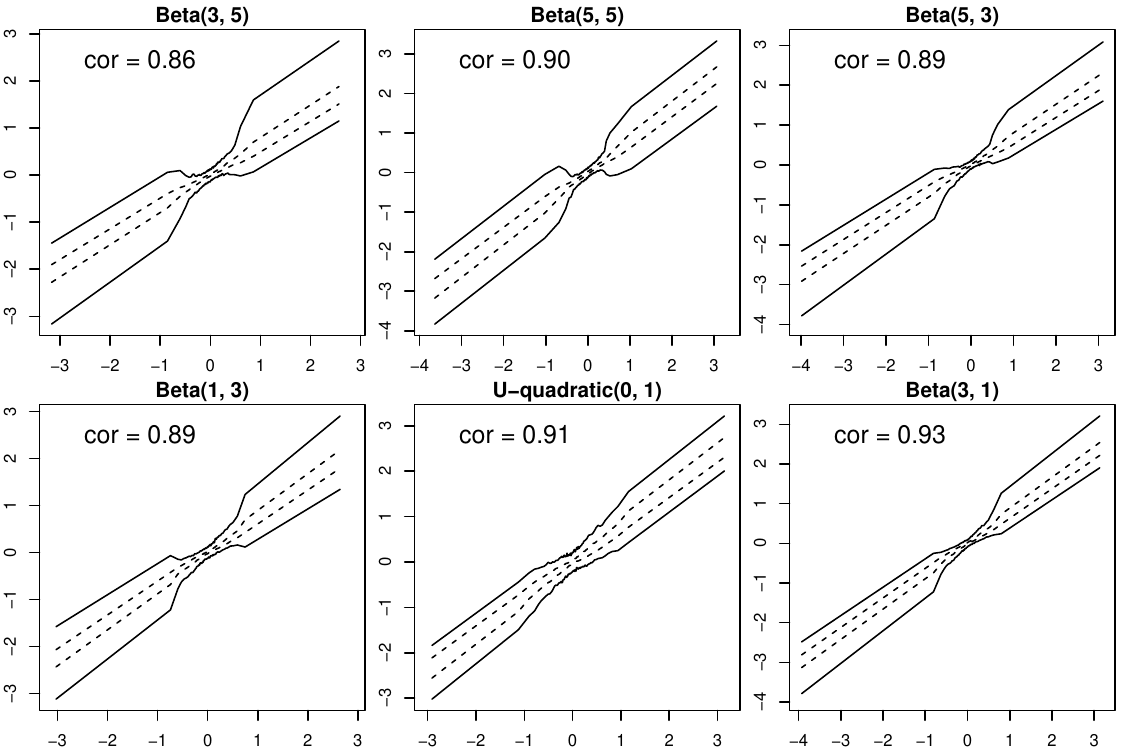"}
	\caption{Prior B-spline coefficients related to general difference penalty and derivative penalty are highly correlated. 50 non-uniform cubic B-splines with a 2nd order penalty are constructed on unevenly spaced knots following different non-uniform distributions. 1000 samples from two priors and plotted against each other. The raw cloud plot with 50000 dots scattered around a straight line is here summarized using 2.5\%, 25\%, 75\% and 97.5\% quantile lines. They imply that 95\% of the dots lie between the two solid lines and 50\% of the dots lie between the two dashed lines.}
	\label{fig7}
\end{figure}

More interesting comparison can be made from the Bayesian perspective \citep{Silverman-some-aspects-of-smoothing-splines} that interprets two penalties as the following Gaussian priors:
\begin{align*}
	\bm{\theta}_{\textrm{gps}} &\sim \textrm{N}(\bm{0}, (\bm{D}_m^{\trans}\bm{D}_m)^-),\\
	\bm{\theta}_{\textrm{os}} &\sim \textrm{N}(\bm{0}, (\bm{K}_m^{\trans}\bm{K}_m)^-),
\end{align*}
where $\bm{M}^-$ denotes the Moore-Penrose generalized inverse of $\bm{M}$. In general, to draw $\bm{\theta} \sim \textrm{N}(\bm{0}, \bm{M}^-)$, we may draw $\bm{e} \sim \textrm{N}(\bm{0}, \bm{I})$ and apply transformation $\bm{\theta} = \bm{\Gamma\Lambda}^{-1/2}\bm{e}$, following the eigendecomposition $\bm{M} = \bm{\Gamma\Lambda\Gamma'}$, where the diagonal matrix $\bm{\Lambda}$ gives nonzero eigenvalues and columns of $\bm{\Gamma}$ give their eigenvectors. Figure \ref{fig6} illustrates 5 samples from two priors and they are surprisingly alike. In this example, 50 non-uniform cubic B-splines with a 2nd order penalty are constructed on unevenly spaced knots following Beta(3, 3) distribution. We have also experimented knots from other non-uniform distributions, finding that samples from two priors are generally very similar and highly correlated. We present such result in Figure \ref{fig7}, where 1000 samples from two priors are plotted against each other. The raw cloud plot showing 50000 dots scattered around a straight line is messy, so we instead summarize it using 2.5\%, 25\%, 75\% and 97.5\% quantile lines. This display also informs us of the percentage of dots between two lines. In all cases, the Pearson correlation between two sets of samples are about 0.9. Given such high similarity between two penalties, or priors, we may expect general P-spline and O-spline to yield similar fit. We now verify this through simulation studies.

\section{Simulation Studies}
\label{section: simulation studies}

In this section, we fit O-spline and standard/general P-splines to simulated $(x_i, y_i)_{1}^{n}$ from $y_i = g(x_i) + \varepsilon_i$, where $g(x)$ is a signal function and $\varepsilon_i$ is a Gaussian white noise, and compare their MSE performance. For all estimators, the number of interior knots $k$, B-spline order $d$ and penalty order $m$ are controlled to be identical. For standard P-spline, we construct uniform B-splines on equidistant knots through the range of $(x_i)_1^n$. For O-spline and general P-spline, we construct non-uniform B-splines on unevenly spaced knots positioned at equal quantiles of $(x_i)_1^n$. To make two knot sequences substantially different, we generate unevenly spaced $(x_i)_1^n$ clustered around $g(x)$'s local extrema. We demonstrate that general P-spline either outperforms O-spline and standard P-spline, or performs equally well.

\begin{figure}
	\centering
	\includegraphics[width = 0.7\columnwidth]{"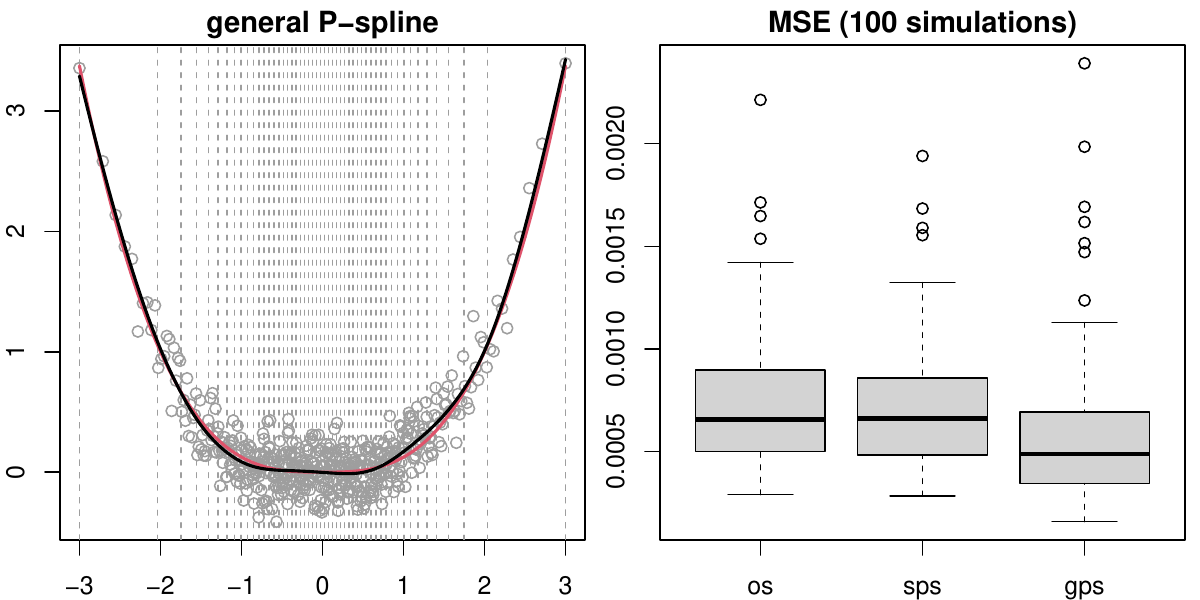"}
	\caption{Left: general P-spline (black) on quantile knots (vertical dashed lines) fitted to noisy data (gray dots) from a U-shaped signal (red). Right: MSE performance of O-spline and standard/general P-splines.}
	\label{fig8}
\end{figure}

\textbf{U-shaped Curve Example} We simulate $n = 500$ noisy observations from $g(x) = \frac{1}{8}|x|^3$, where $(x_i)_1^n$ are drawn from $\textrm{N}(0, 1)$ distribution, and fit a cubic spline ($d = 4$) represented by B-splines on $k = 50$ interior knots with a 2nd order penalty ($m = 2$). In Figure \ref{fig8}, the boxplot of MSE based on 100 simulations suggests that general P-spline outperforms O-spline and standard P-spline.

\begin{figure}
	\centering
	\includegraphics[width = 0.7\columnwidth]{"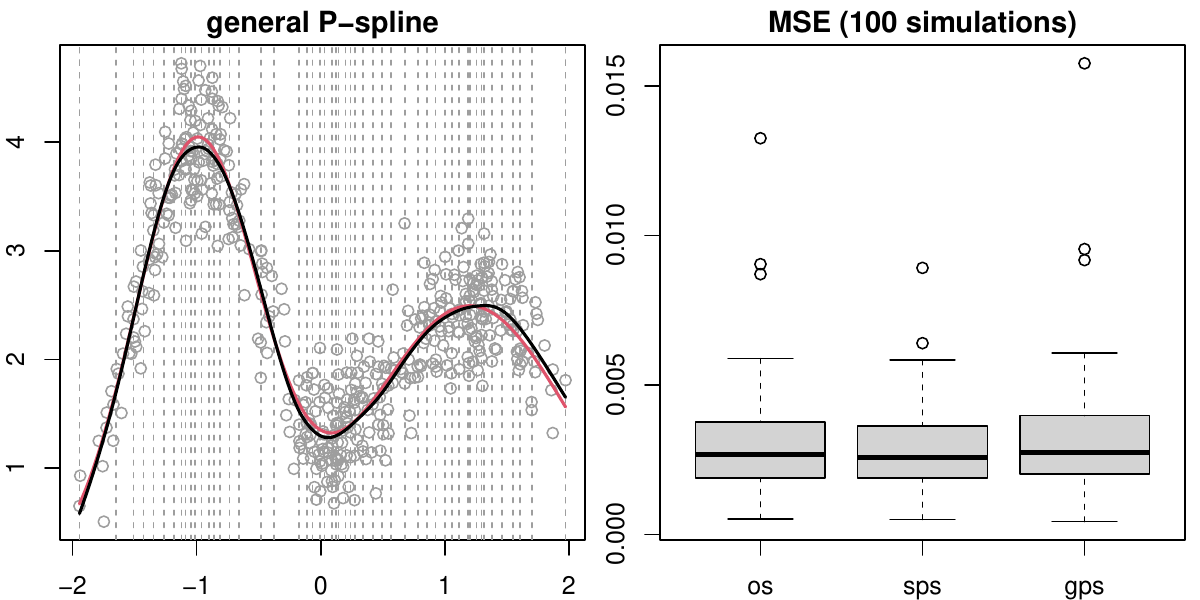"}
	\caption{Left: general P-spline (black) on quantile knots (vertical dashed lines) fitted to noisy data (gray dots) from a normal mixture signal (red). Right: MSE performance of O-spline and standard/general P-splines.}
	\label{fig9}
\end{figure}

\textbf{Normal Mixture Example} We simulate $n = 500$ noisy observations from $g(x) = 5f_1(x) + 5f_2(x)$, where $f_1(x)$ and $f_2(x)$ are density functions of $\textrm{N}(-1, 0.5^2)$ and $\textrm{N}(1.2, 0.8^2)$ distributions, and $(x_i)_1^n$ are drawn from a normal mixture $\frac{1}{3}\textrm{N}(-1, 0.35^2) + \frac{1}{3}\textrm{N}(1.2, 0.35^2) + \frac{1}{3}\textrm{N}(0.1, 0.2^2)$. We fit a cubic spline ($d = 4$) represented by B-splines on $k = 50$ interior knots with a 2nd order penalty ($m = 2$). In Figrue \ref{fig9}, the boxplot of MSE based on 100 simulations suggests that O-spline and standard/general P-splines perform equally well.

\begin{figure}
	\centering
	\includegraphics[width = 0.7\columnwidth]{"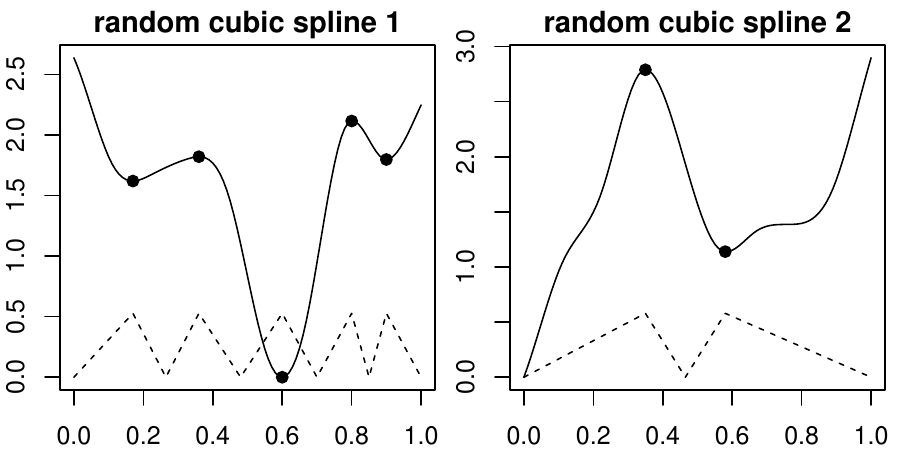"}
	\caption{Two examples of random cubic spline (solid) on domain [0, 1]. Unevenly spaced $(x_i)_1^n$ are sampled from a ``tent'' distribution whose density (dashed) peaks at local extrema (dots) of the spline.}
	\label{fig10}
\end{figure}

\textbf{Random Curve Example} We now design a simulation study that is based on a family of random curves rather than a specific $g(x)$. To be precise, we repeat the following procedure for $l = 1, 2, \ldots, N$:
\begin{enumerate}
	\item generate a random curve $g^{\iter{l}}(x)$ on $[0, 1]$;
	\item evaluate $g_i^{\iter{l}} = g^{\iter{l}}(x_i^{\iter{l}})$ at unevenly spaced $(x_i^{\iter{l}})_1^n$;
	\item simulate noisy observations $y_i^{\iter{l}} = g_i^{\iter{l}} + \varepsilon_i^{\iter{l}}$, with $\varepsilon_i^{\iter{l}} \iid \textrm{N}(g_i^{\iter{l}}, \sigma_{\sub{l}}^2)$;
	\item estimate $\hat{f}_i^{\iter{l}} = \hat{f}^{\iter{l}}(x_i^{\iter{l}})$ from $(x_i^{\iter{l}}, y_i^{\iter{l}})$ using O-spline and standard/general P-splines;
	\item compute relative MSE: $\delta_{\sub{l}} = \frac{1}{n}\sum_{i = 1}^{n}(\hat{f}_i^{\iter{l}} - g_i^{\iter{l}})^2/\sigma_{\sub{l}}^2$, for each fitted spline.
\end{enumerate}
At step 1, we construct $g^{\iter{l}}(x)$ as an order-$d$ spline with $p^{\star} = 3d$ uniform B-splines on $4d$ equidistant knots. To make $g(x)$ smooth, we take B-spline coefficients to be a 1st order random walk (which is the prior related to the 1st order standard difference penalty). See Figure \ref{fig10} for two example random cubic splines ($d = 4$) constructed this way. We can express $g^{\iter{l}}(x)$ using more B-splines, but it produces more local extrema and require more observations (a bigger $n$) for a reasonable estimation. At step 2, we sample $(x_i^{\iter{l}})_1^n$ from some ``tent'' distribution whose density is illustrated by the dashed line in Figure \ref{fig10}. At step 3, we specify $\sigma_{\sub{l}} = \gamma\times\textrm{s.e.}\{(g_i^{\iter{l}})_1^n\}$, where $\textrm{s.e.}\{(g_i^{\iter{l}})_1^n\}$ is the standard error of signal and $\gamma$ is a \textit{noise-to-signal ratio}. We hold $\gamma$ fixed through $l = 1, 2, \ldots, N$, but $\sigma_{\sub{l}}$ will vary from signal to signal (which explains why in step 5 we divide raw MSE by $\sigma_{\sub{l}}^2$). At step 4, we represent penalized B-splines estimator $f^{\iter{l}}(x)$ using $p = 5p^{\star} = 15d$ B-splines constructed on $k = p - d = 14d$ interior knots. There is no need to use a larger $p$ or $k$. The choice here is already more than adequate, because by the construction in step 1, the true degree of freedom in $g^{\iter{l}}(x)$ is no bigger than $p^{\star}$, thus $p^{\star} - p = 12d$ or more B-spline coefficients in $f^{\iter{l}}(x)$ will be suppressed by the wiggliness penalty.

\begin{figure*}[ht]
	\centering
	\includegraphics[width = \columnwidth]{"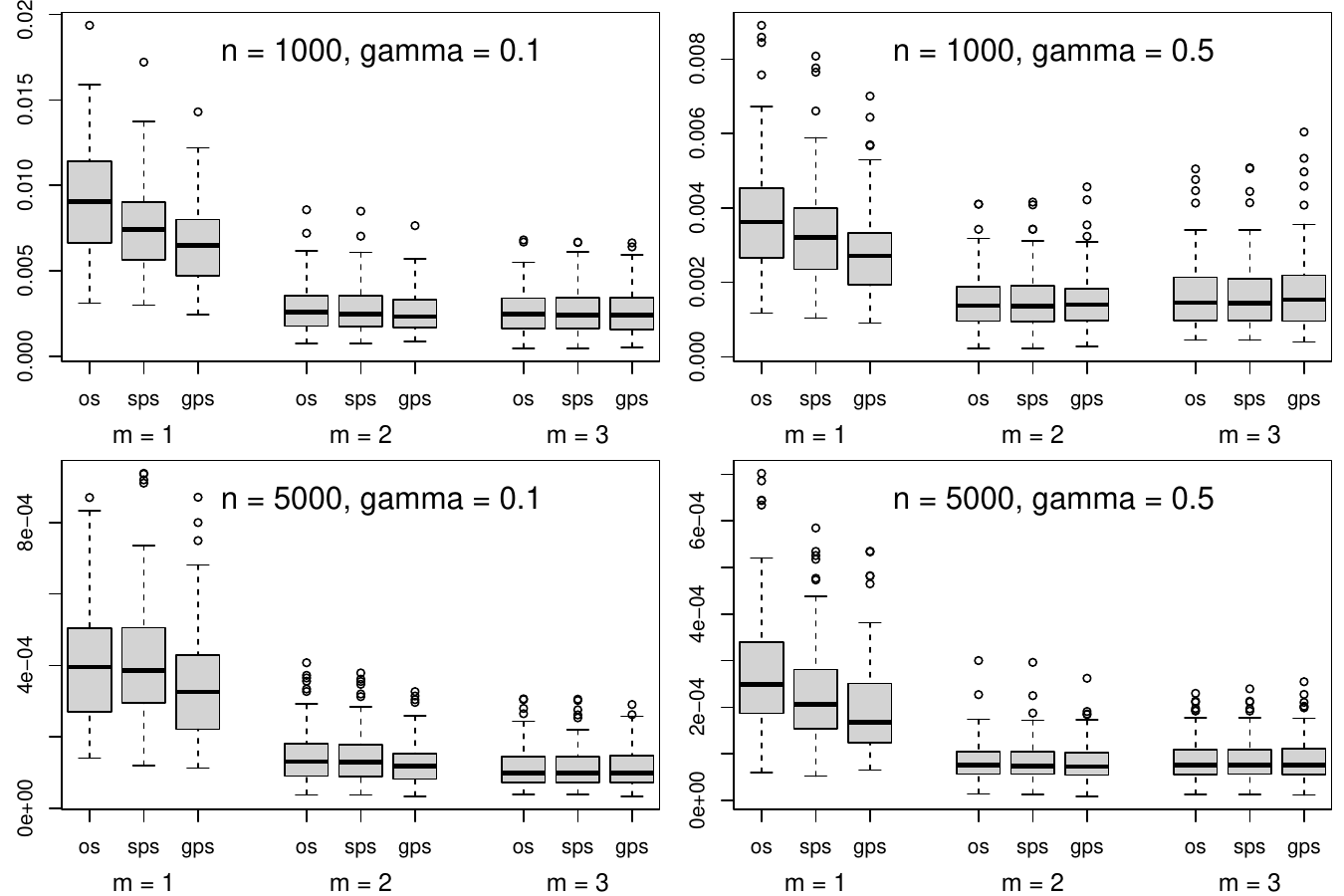"}
	\caption{MSE performance of O-spline and standard/general P-splines for cubic spline ($d = 4$) with various choices of $n$ (sample size) and $\gamma$ (noise-to-signal ratio) under $N = 100$ simulations.}
	\label{fig11}
\end{figure*}

To start the simulation we need to specify the number of simulations $N$, B-spline order $d$, number of observations $n$ and noise-to-signal ratio $\gamma$. For penalty order, we attempt all $m = 1, 2, \ldots, d - 1$. Figure \ref{fig11} shows the simulation result for cubic splines ($d = 4$) with $N = 100$ and four different $(n, \gamma)$ pairs: (1000, 0.1), (1000, 0.5), (5000, 0.1) and (5000, 0.5). We see that general P-spline outperforms O-spline and standard P-spline when $m = 1$ and performs equally well otherwise.

\clearpage
\section{Real Data Examples}
\label{section: real data examples}

Our simulation studies are composed of well-designed examples, in which (1) the underlying smooth function $g(x)$ has spatially homogeneous smoothness; (2) sampling locations $(x_i)_1^n$ are properly distributed so that the curvature of $g(x)$ can be well estimated; (3) interior knots are well spread over $(x_i)_1^n$; (4) sufficient number of interior knots are positioned to eliminate sensitivity in exact knot locations. Only when all these conditions are met, is general P-spline competitive with standard P-spline in MSE performance. Unfortunately, this may not be the case when smoothing real-world data. In particular, the knot placement issue related to (3) and (4) is by and large a subjective matter to be decided by practitioners. As a result, general P-spline on unevenly spaced knots may produce a more satisfactory fit than standard P-spline on equidistant knots for some datasets, and vise versa, depending on the knot sequence being used.

In this section, we demonstrate the impact of knot placement on penalized B-splines, by comparing the quality of standard/general P-splines fit to real datasets. To narrow our discussion, we focus on a popular type of unevenly spaced knots, called quantile knots \citep{SemiPar-book-2003}, that are positioned at equal quantiles of $(x_i)_1^n$. This leaves the number of interior knots, $k$, the only factor that controls equidistant knots and quantile knots. We will see that for small to moderate $k$, the BMC longitudinal data favor general P-spline on quantile knots, while the fossil shell data favor standard P-spline on equidistant knots. We do not interpret this phenomenon as one type of knots being superior than the other. Rather, we would say that for a particular dataset, one type of knots is more \textit{effective} than the other when $k$ is restricted in size. After all, as $k$ becomes adequately big, exact knot locations no longer matter.

\subsection{BMC Longitudinal Data}
\label{subsection: BMC longitudinal data}

The Pediatric Bone Mineral Accrual Study (PBMAS) \citep{PBMAS-study-original,PBMAS-study} was launched in 1991 by the University of Saskatchewan, aiming to investigate the accumulation of bone mineral content (BMC) in growing children. At first, 109 boys and 113 girls aged 8 $\sim$ 15 from two elementary schools in Saskatoon (the largest city in Saskatchewan, Canada) were involved for annual scan. 31 new subjects were recruited in 1992 $\sim$ 1993. Over the years, the study was interrupted several times, with data collection in 1991 $\sim$ 1997, 2003 $\sim$ 2005, 2007 $\sim$ 2011 and 2016 $\sim$ 2017. Many subjects dropped out as they left school, but some managed to return for follow-up scans as young adults. As of 2017, the dataset has 982 and 1175 measurements from males and females, respectively.

\begin{figure}
	\includegraphics[width = \columnwidth]{"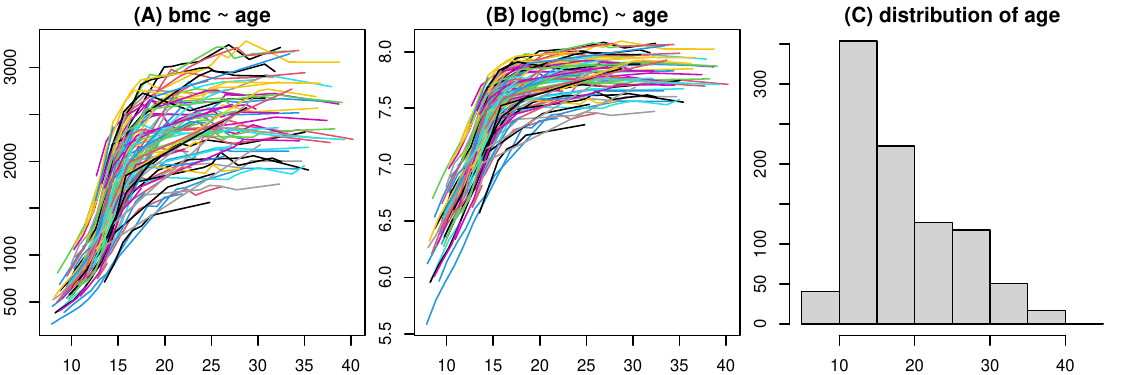"}
	\caption{(A) 932 BMC measurements (in grams) for 112 white ethnic males. (B) log-transformed measurements. (C) distribution of age when measurement was taken.}
	\label{fig12}
\end{figure}

\cite{BMC-trajectory-modeling} studied white ethnic subjects in the dataset (comprising 112 males and 127 females) with a linear mixed-effects model \citep{linear-mixed-models-book} for either gender group as follows:
\begin{equation*}
	\textrm{bmc}_{j, \textrm{age}} = \sum_{u = 1}^{p_1}B_u^{\iter{1}}(\textrm{age})\beta_u + \sum_{v = 1}^{p_2}B_v^{\iter{2}}(\textrm{age})\alpha_{j,v} + e_{j, \textrm{age}},
\end{equation*}
where $(B_u^{\iter{1}}(\textrm{age}))_{1}^{p_1}$ and $(B_v^{\iter{2}}(\textrm{age}))_{1}^{p_2}$ are cubic B-splines with fixed-effect coefficients $(\beta_u)_{1}^{p1}$ and random-effect coefficients $(\beta_v)_{1}^{p2}$, respectively, and $e_{j, \textrm{age}} \iid \textrm{N}(0, \sigma_e^2)$ is a measurement error. The first summation models the population BMC trajectory, and the second models the deviation of subject $j$'s BMC trajectory from the former. When fitting this model using standard \textbf{R} packages for linear mixed-effects models, namely \textbf{nlme} \citep{nlme-package} and \textbf{lme4} \citep{lme4-paper, lme4-package}, it is problematic to estimate the covariance matrix of random effects, because either the estimation algorithm fails to converge, or the estimated covariance matrix is singular. This is not resolved until $p_2$ is as small as 4. As a result, complexity of subject trajectories allowed by this model is very restricted.

A better strategy is to consider an additive mixed model \citep{highland-GAMMs-book, hierarchical-GAM} as follows:
\begin{equation*}
	\log(\textrm{bmc})_{j, \textrm{age}} = f(\textrm{age}) + f_j(\textrm{age}) + e_{j, \textrm{age}},
\end{equation*}
where $f$ (a smooth fixed-effect function) is the population log(BMC) trajectory, $f_j$ (a smooth random-effect function) is the deviation of subject $j$'s log(BMC) trajectory from $f$ and $e_{j, \textrm{age}} \iid \textrm{N}(0, \sigma_e^2)$ is a measurement error. Constructing $f$ and $f_j$ as penalized B-splines, the model can be estimated in the framework of generalized additive models (GAMs) \citep{Wood-GAMs-book} using \textbf{R} package \textbf{mgcv} \citep{mgcv-package}. Note that we choose to model log(BMC) instead of BMC, because Figure \ref{fig12} shows that BMC does not have constant variance while log(BMC) does.

\begin{figure*}[t]
	\centering
	\includegraphics[width = 0.95\columnwidth]{"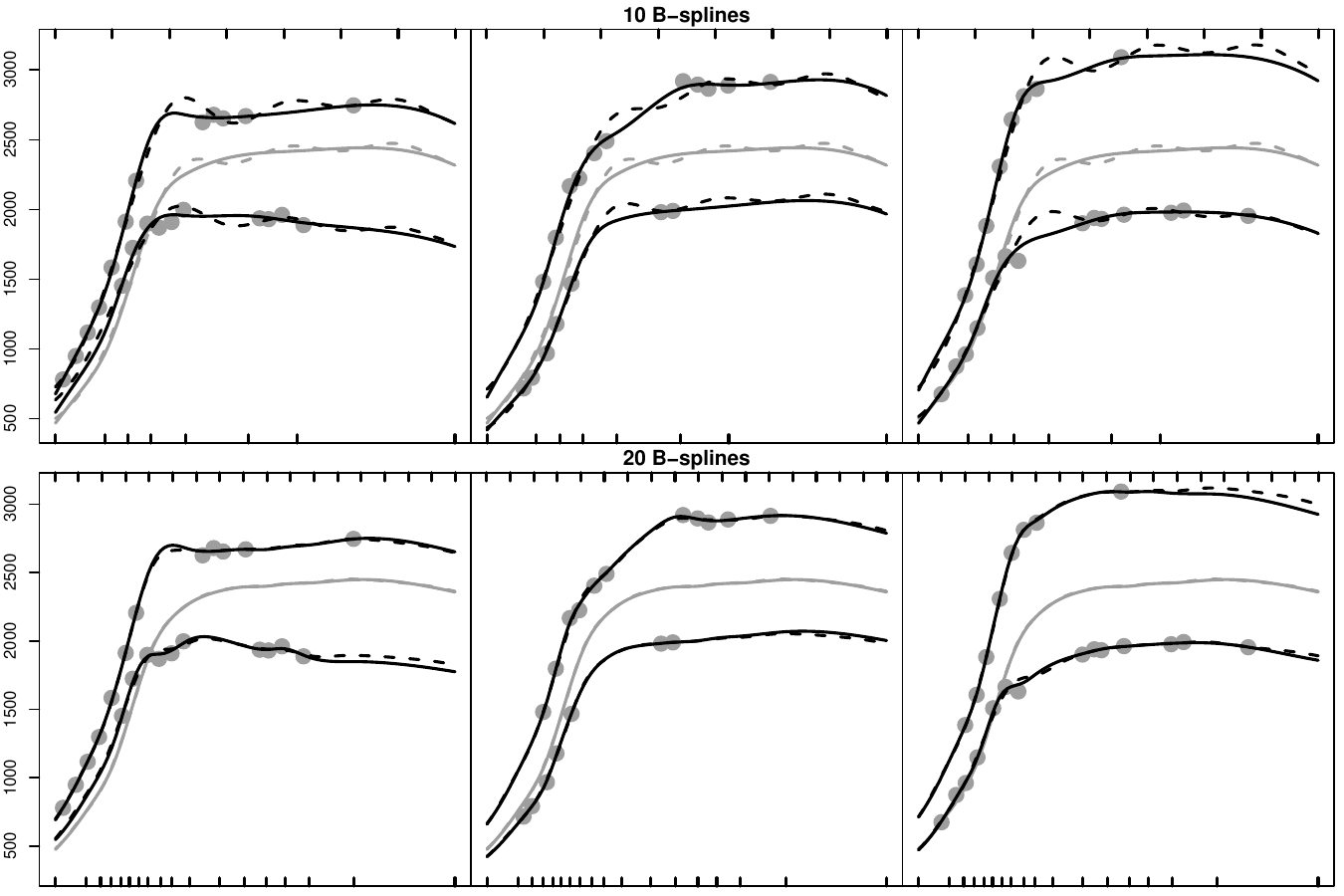"}
	\caption{Estimated BMC trajectories with general P-spline (solid) and standard P-spline (dashed) when $p = $ 10 (top row) and 20 (bottom row). Gray curves: population trajectories; black curves: subject trajectories for six selected subjects; gray dots: data for the selected subjects; rugs on the bottom axis: quantile knots for general P-spline; rugs on the top axis: equidistant knots for standard P-spline.}
	\label{fig13}
\end{figure*}

\begin{figure}
	\centering
	\includegraphics[width = 0.7\columnwidth]{"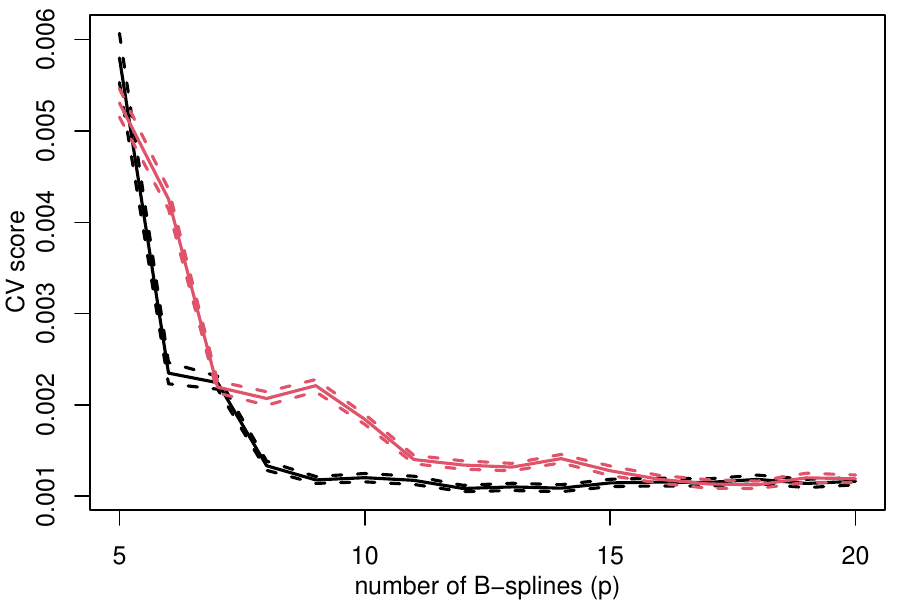"}
	\caption{Cross-validation score (solid) and its 2 standard error bounds (dashed) for general P-spline (black) and standard P-spline (red) at $p = 5, 6, \ldots, 20$. For BMC data, general P-spline on quantile knots is more effective than standard P-spline on equidistant knots.}
	\label{fig14}
\end{figure}

We represent both $f$ and $f_j$ by cubic B-splines. For penalty order, we choose $m = 2$ for $f$, but $m = 1$ for $f_j$. This subtlety comes from the fact that fewer subjects came back for measurements as they grew older. For example, the number of subjects (out of 112 in total) with no observations after the age of 20, 25 and 30 are 36, 50 and 60, respectively. Using $m = 2$ for $f_j$ includes a random slope in the penalty's null space that would lead to linear extrapolation for these subjects. But Figure \ref{fig12} shows that log(BMC) trajectories are relatively flat after the age of 20. Thus, it is more reasonable to exclude this random slope by choosing $m = 1$, leaving only a random intercept in the penalty's null space.

We set up both $f$ and $f_j$ using $p$ cubic B-splines on the same knot sequence, which comprises $k = p - 4$ interior knots and other necessary boundary knots. Note that there are more measurements at early ages (see Figure \ref{fig12}), so quantile knots are right-skewed. Figure \ref{fig13} shows that when $p = 10$, standard P-spline fit is suspiciously wiggly, while general P-spline fit is plausibly smooth. But the difference between two P-spline fits is much less noticeable when $p = 20$. The message is that for BMC data, general P-spline on quantile knots is more effective than standard P-spline on equidistant knots. Let's demonstrate this in a more quantitative way by computing the 10-fold cross-validation score for $p = 5, 6, \ldots, 20$. Figure \ref{fig14} shows that the score for general P-spline decreases with $p$ faster. It already achieves a satisfactory fit when $p = 9$, but standard P-spline fit requires $p = 17$ to attain about the same quality.

\subsection{Fossil Shell Data}

The fossil shell data (available in \textbf{R} package \textbf{SemiPar} \citep{SemiPar-package}) raised a debate \citep{SemiPar-article-2009, SemiPar-with-O-splines, P-splines-20-years} on the superiority of either equidistant knots or quantile knots. We revisit this example, and aligned with these works, consider an estimate by \textbf{R} function \texttt{smooth.spline} as ``truth''. Note that previous demonstration with quantile knots was via O-spline. We now replace it by general P-spline.

\begin{figure}
	\centering
	\includegraphics[width = 0.8\columnwidth]{"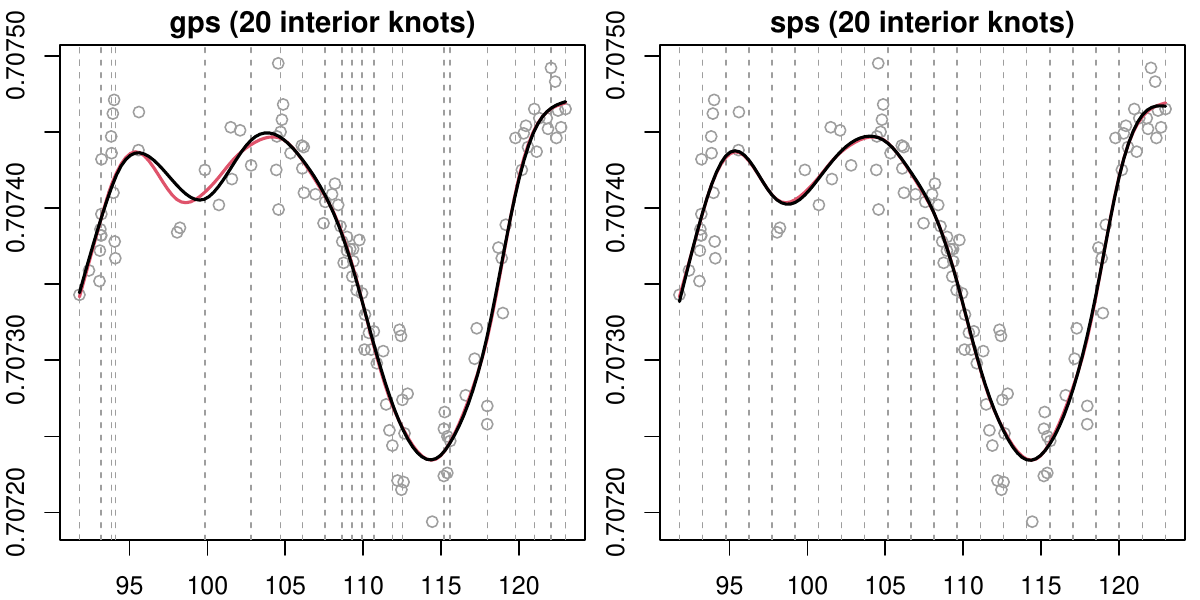"}
	\caption{Standard/general P-spline fits (black) and \texttt{smooth.spline} fit (red) for fossil shell data (gray dots). Vertical dashed lines: knot locations for standard/general P-splines.}
	\label{fig15}
\end{figure}

\begin{figure}
	\centering
	\includegraphics[width = 0.8\columnwidth]{"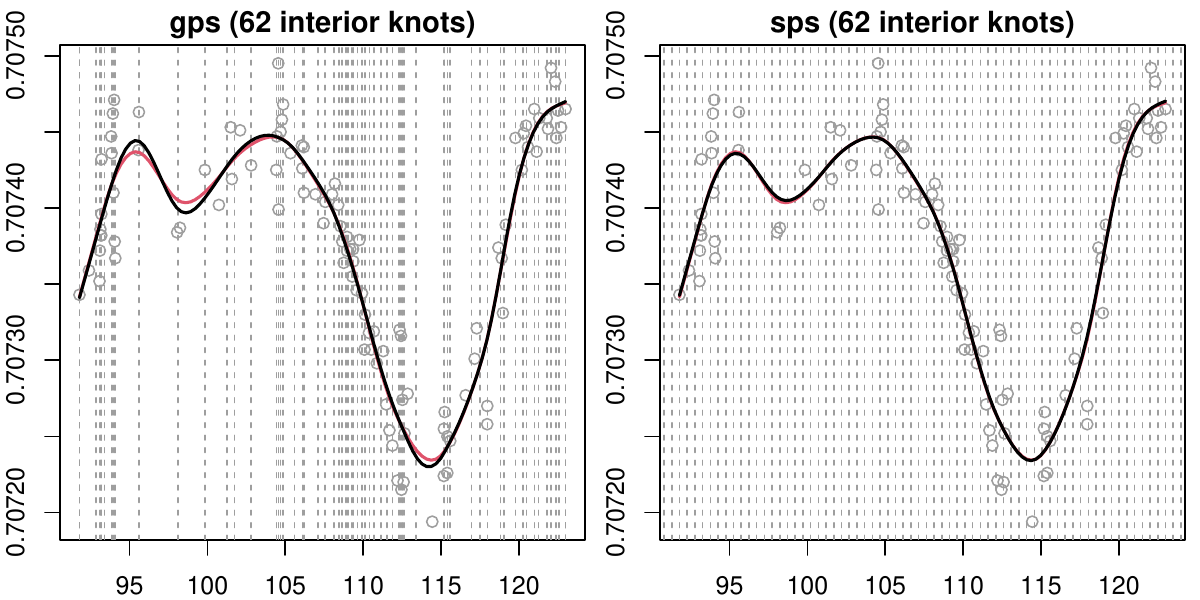"}
	\caption{Standard/general P-spline fits (black) and \texttt{smooth.spline} fit (red) for fossil shell data (gray dots). Vertical dashed lines: knot locations for standard/general P-splines.}
	\label{fig16}
\end{figure}

\cite{P-splines-20-years} claimed that equidistant knots are superior, because Figure \ref{fig15} shows that general P-spline on 20 quantile knots fails to capture the local minimum near $x = 98$ correctly, while standard P-spline on 20 equidistant knots has no trouble with this. They criticized that placing quantile knots is ``the root of all evil''. However, they did not realize that \texttt{smooth.spline} fits an O-spline on quantile knots, too! In fact, Figure \ref{fig16} shows that re-estimating general P-spline on the same 62 quantile knots positioned by \texttt{smooth.spline} seems to outperform both O-spline and standard P-spline, with better approximation to data near $x$ = 96, 98 and 114 without sacrificing any smoothness. Since the true function is actually not known, we report residual sum of squares (RSS) rather than MSE for these estimates, which are $\textrm{RSS}_{\textrm{sps}} = 5.87 \times 10 ^ {-8}$, $\textrm{RSS}_{\textrm{os}} = 5.78 \times 10 ^ {-8}$ and $\textrm{RSS}_{\textrm{gps}} = 5.74 \times 10 ^ {-8}$. Clearly, with sufficient number of knots, general P-spline attains the best fit.

\begin{figure}
	\centering
	\includegraphics[width = 0.7\columnwidth]{"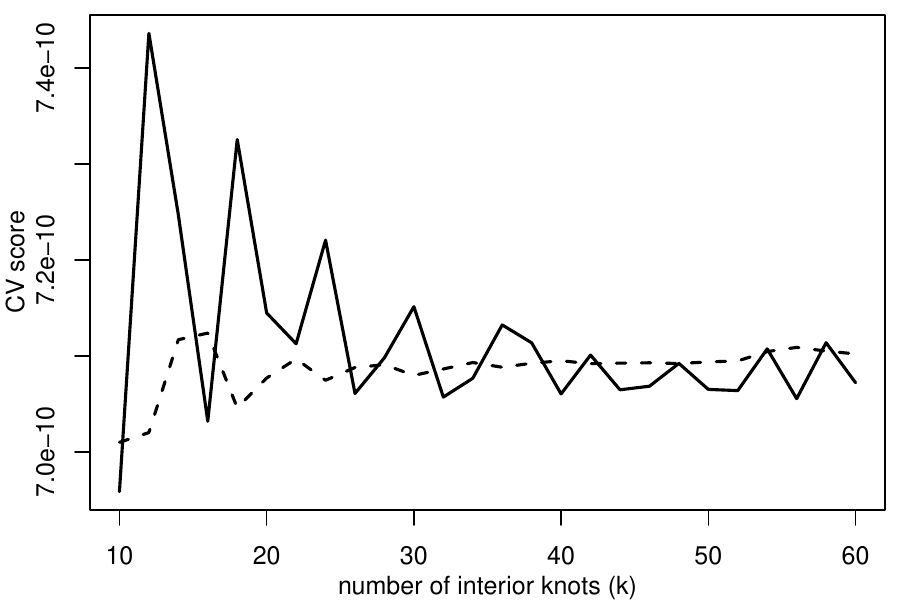"}
	\caption{Leave-one-out generalized cross-validation scores of general P-spline (solid) and standard P-spline (dashed) for $k = 10, 11, \ldots, 60$. For fossil shell data, standard P-spline on equidistant knots is more effective than general P-spline on quantile knots.}
	\label{fig17}
\end{figure}

\begin{figure}
	\centering
	\includegraphics[width = \columnwidth]{"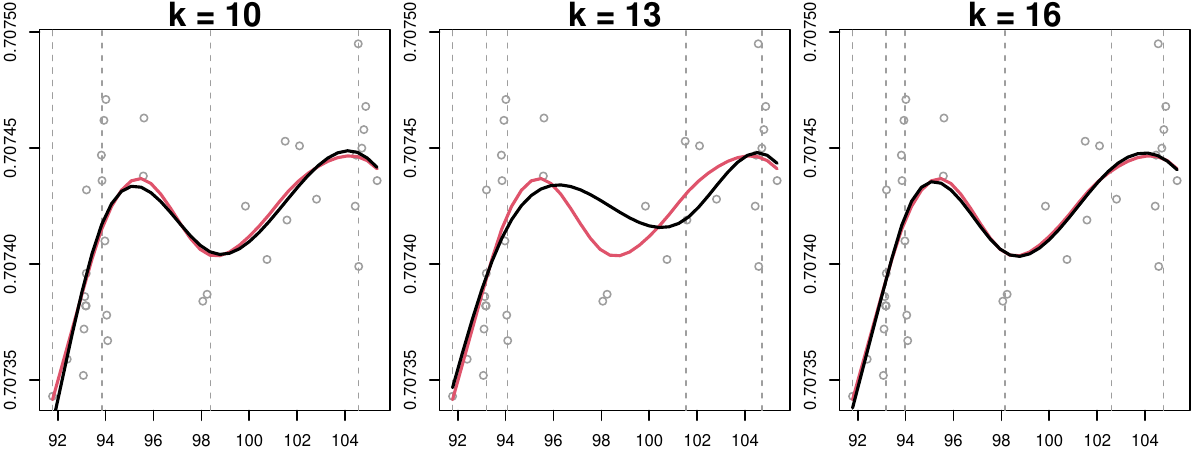"}
	\caption{A zoomed-in display of general P-spline fit (black) and \texttt{smooth.spline} fit (red) on $[92, 106]$. A small $k$ may unfortunately miss the critical knot around $x = 98$.}
	\label{fig18}
\end{figure}

Again, we argue that which type of knots appears more advantageous is conditional on $k$ (number of interior knots). Figure \ref{fig17} shows the cross-validation score of standard/general P-splines for $k = 10, 11, \ldots, 60$. It is astounding to spot a zigzag in general P-spline's score. Figure \ref{fig18} shows that for fossil shell data, general P-spline on quantile knots is very sensitive to the existence of a knot around $x = 98$, which a small $k$ may unfortunately miss. Such sensitivity gets reduced as $k$ increases, and after $k = 40$, general P-spline begins to outperform standard P-spline by a small margin. But before that, standard P-spline almost always gains the upper hand. Hence, for small to moderate $k$, standard P-spline is more effective for fossil shell data.

\clearpage
\section{Discussion}

We proposed a new penalized B-splines estimator, the general P-spline, to accommodate B-splines on arbitrary knot sequences, like non-uniform B-splines on unevenly spaced knots. It is a complement to Eilers and Marx's standard P-spline that only makes sense for uniform B-splines on equidistant knots. At its core, we derived a novel general difference matrix $\bm{D}_m$ that accounts for uneven knot spacing, to replace the standard difference matrix $\bm{\Delta}_m$ tailored for equidistant knots. Both P-spline variants are practically useful, because either one may produce a more satisfactory fit than the other depending on the knot sequence being used, in addition to the data being analyzed. Therefore, practitioners should try out both before betting on either one.

General P-spline is closely related to O-spline. The sandwich formula $\bm{S}_m = \bm{D}_m^{\trans}\bm{\bar{S}}_m\bm{D}_m$ explicitly links general difference penalty and derivative penalty together. On one hand, both penalties have the same null space. On the other hand, there is a one-to-one correspondence between two penalty functions. When interpreting both penalties as Gaussian priors for B-spline coefficients, samples from two prior distributions are highly alike and correlated. Given all these similarities, they are expected to be equally powerful for wiggliness control. In fact, simulations shows that general P-spline either outperforms O-spline in terms of MSE, or performs equally well. Thus, we believe general P-spline to be a superior replacement of O-spline.

The general difference penalty can be modified to be even more similar to derivative penalty, by an extra weighting process: $\bm{\tilde{D}}_m = \bm{B}_m^{1/2}\bm{D}_m$, where $\bm{B}_m$ is a diagonal matrix containing $\bm{\bar{S}}_m$'s diagonal elements, i.e., the $j$\textsuperscript{th} diagonal entry in $\bm{B}_m$ is $\int_{a}^{b}B_{j, d - m}(x)^2\textrm{d}x$. In this way, $\bm{\tilde{D}}_m$ also accounts for non-uniform B-splines' inhomogeneous shapes. However, it makes the penalty difficult to interpret. Without this extra weighting, the general difference penalty $\textrm{PEN}_{\textrm{gps}}^{\ord{m}}(\bm{\beta}) = \|\bm{D}_m\bm{\beta}\|^2 = \|\bm{\beta}_m\|^2$ stands for the squared L\textsubscript{2} norm of $f^{\ord{m}}(x)$'s B-spline coefficients. With this extra weighting, it is the the squared L\textsubscript{2} norm of $f^{\ord{m}}(x)$'s scaled B-spline coefficients, where the $j$\textsuperscript{th} coefficient is multiplied by the square root of the area under the $j$\textsuperscript{th} B-spline curve! We do not appreciate this idea much, because it is not our intention to approximate derivative penalty as best as we can. In addition, computation of general difference penalty loses its simplicity as it requires numerical integration to work out $\bm{B}_m$.

The sandwich formula itself is also useful, enabling us to compute a sparse ``root'' of the derivative penalty matrix $\bm{S}_m$ for the first time. We denoted this matrix by $\bm{K}_m$ when elaborating the Bayesian interpretation of the derivative penalty $\textrm{PEN}_{\textrm{os}}^{\ord{m}}(\bm{\beta}) = \|\bm{K}_m\bm{\beta}\|^2$, but have not shown what it is like. Consider the example knot sequence used in Section \ref{subsection: introducing general P-spline}. For cubic B-splines ($d = 4$), it can be computed that:
\begin{gather*}
	\bm{K}_1 = \begin{bmatrix}
		-1.34 & \phantom{-}1.07 & \phantom{-}0.05 & \phantom{-}0.22\\
		& -0.57 & \phantom{-}0.41 & \phantom{-}0.14 & \phantom{-}0.01\\
		& & -0.62 & \phantom{-}0.38 & -0.25 & 0.48\\
		& & &-0.59 & \phantom{-}0.16 & 0.43\\
		& & & & -1.18 & 1.18
	\end{bmatrix},\\
	\bm{K}_2 = \begin{bmatrix}
		\phantom{-}3.46 & -4.43 & \phantom{-}0.82 & \phantom{-}0.14\\
		& \phantom{-}0.64 & -0.94 & \phantom{-}0.07 & \phantom{-}0.23\\  
		& & \phantom{-}0.47 & -0.74 & -0.80 & 1.07\\
		& & & \phantom{-}1.10 & -4.39 & 3.30
	\end{bmatrix},\\
	\bm{K}_3 = \begin{bmatrix}
		-6.00 & \phantom{-}8.67 & -3.17 & \phantom{-}0.50\\
		& -0.47 & \phantom{-}1.18 & -1.18 & 0.47\\
		& & -0.50 & \phantom{-}3.17 & -8.67 & 6.00
	\end{bmatrix}.
\end{gather*}
Previously, to find $\bm{K}_m$ such that $\bm{S}_m = \bm{K}_m^{\trans}\bm{K}_m$, we have to resort to $\bm{S}_m$'s eigendecomposition due to its rank deficiency. And consequently, sparsity is destroyed and we get, for example,
\begin{equation*}
	\bm{K}_2 = \begin{bmatrix}
		-2.43 & \phantom{-}3.17 & -0.74 & -0.74 & \phantom{-}3.17 & -2.43\\
		\phantom{-}2.44 & -3.15 & \phantom{-}0.57 & -0.57 & \phantom{-}3.15 & -2.44\\
		-0.25 & -0.04 & \phantom{-}0.85 & -0.85 & \phantom{-}0.04 & \phantom{-}0.25\\
		\phantom{-}0.31 & \phantom{-}0.13 & -0.44 & -0.44 & \phantom{-}0.13 & \phantom{-}0.31
	\end{bmatrix}.
\end{equation*}
Why would a sparse ``root'' be interesting? Because it is key to computational efficiency, if we are to adapt the above L\textsubscript{2} penalty to L\textsubscript{1} penalty $\|\bm{K}_m\bm{\beta}\|_1$ for locally adaptive regression \citep{locally-adaptive-regression-splines,genlasso-dual-path-theory, genlasso-dual-path-implementation,polynomial-trendfilter, trendfilter-ADMM,L1-P-splines}.

General P-spline is motivated by the use of irregularly spaced knots. For demonstration, the popular quantile knots that facilitate automatic knot placement given the number of interior knots ($k$) are used in our simulation studies and real data examples. In many literature, quantile knots are also considered ``evenly spaced'', because it results in about identical number of $x$-values between nearby knots. Therefore, it is quite a natural choice when smoothing non-uniformly distributed data. But as the fossil shell data example shows, when $k$ is not sufficiently big, quantile knots may miss some critical location for curvature estimation of the function. \cite{YaoFang-knot-placement} proposed a two-step approach to improve such knot placement. Basically, after fitting a general P-spline on $k$ quantile knots, we augment the knot sequence by adding locations of all local extrema of the fitted spline and refit a general P-spline. The idea is simple, but effective in practice.

\begin{figure}
	\includegraphics[width = \columnwidth]{"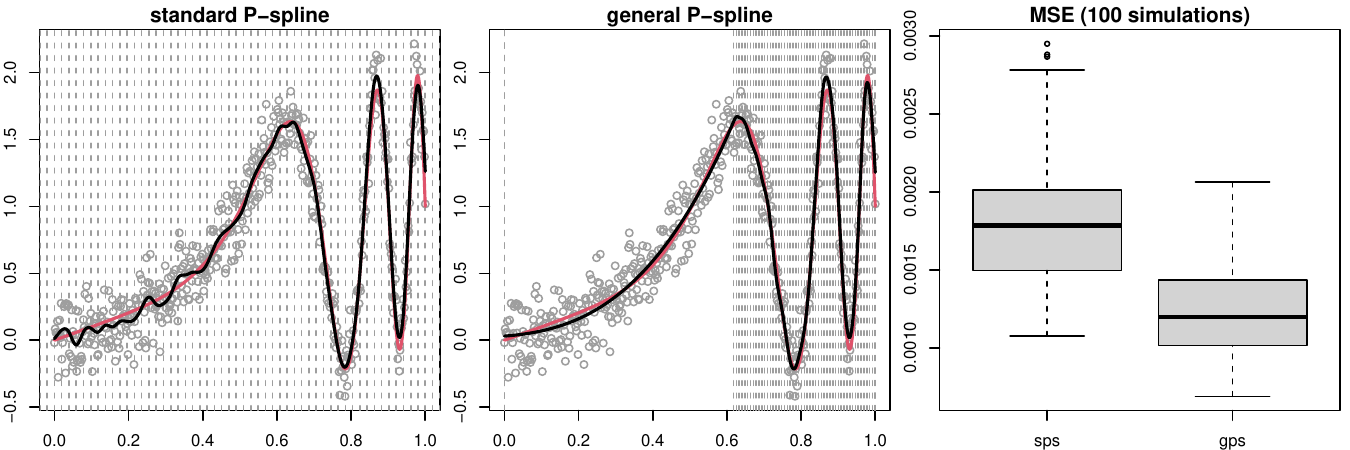"}
	\caption{Unevenly spaced knots are necessary for satisfactory estimation of a spatially inhomogeneous function. A crude yet adaptive specification of a knots superset is sufficient, and the penalty will suppress excessive amount of degree of freedom to prevent overfitting.}
	\label{fig19}
\end{figure}

In full generality, knot locations can be optimized via sophisticated and computationally intensive knot selection or ``free-knot'' estimation \citep{approximation-to-data-by-splines-with-free-knots,penalized-free-knot-splines,bayesian-free-knot-splines,adaptive-free-knot-splines,free-knot-smoothing-functional-data,freeknotsplines-archived-by-CRAN}. This idea is most relevant to regression splines with no wiggliness penalty, in which case a fitted spline is extremely sensitive to exact knot locations. In penalized regression, however, there is no need to perform such an exhaustive and refined search. In principle, a crude guess at a knots superset would be sufficient, as the penalty will suppress excessive amount of degree of freedom to prevent overfitting. In this regard, standard P-spline always makes a blind guess, taking this superset to be equidistant knots whatever the underlying function is, whereas general P-spline encourages adapting this superset if any prior information of the function is available. This is particularly important when estimating a function with spatially inhomogeneous smoothness, because standard P-spline would never approximate this function uniformly, no matter how many equidistant knots are positioned. For example, consider smoothing 500 noisy observations of $g(x) = x + \sin(5\pi x^5)$, $x \in [0, 1]$ at equidistant $x$-values. Figure \ref{fig19} shows that a standard P-spline on 50 equidistant knots ends up with a wiggly fit on $g(x)$'s ``tail''. Fitting a general P-spline on quantile knots does not resolve this, because quantile knots are also equidistant in this case. Now let's relocate these 50 knots, with no interior knot between 0 and about 0.6, and equidistant knots in the rest of the domain. We do so because the scatterplot convinces us that a single cubic polynomial segment should be adequate for modeling the ``tail''. Figure \ref{fig19} shows that the general P-spline fit on this knot sequence is a success, and it outperforms standard P-spline fit by a large margin in terms of MSE. Of course, this is only a toy example of a crude yet adaptive specification of a knots superset. How to automate such heuristic for an arbitrary problem is worth further research.

\section*{Appendix: Spline and B-Splines}

An order-$d$ spline $f(x)$ defined on domain $[a, b]$ comprises smoothly connected polynomial segments of degree $d - 1$, with $d - 2$ continuous derivatives at their interior knots (or break points) $a < s_1 < s_2 < \ldots < s_k < b$. Equivalently, it may be expressed as a linear combination of $p = k + d$ order-$d$ B-splines, whose construction requires $d - 1$ arbitrarily positioned auxiliary boundary knots on each side of $[a, b]$. In total, there are $K = k + 2d$ knots (including $a$ and $b$), denoted by $(t_j)_{1}^{K}$.

\begin{figure}[h]
	\centering
	\includegraphics[width = 0.577\columnwidth]{"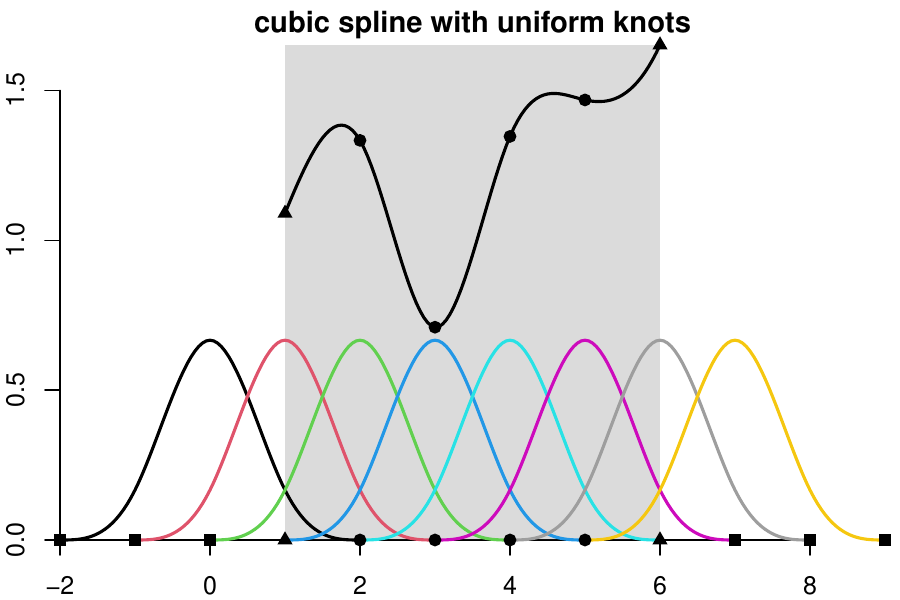"}
	\caption{B-splines (bottom) representation of a cubic spline (top). Shaded area: domain; triangles: domain endpoints; circles: interior knots (break points of polynomial pieces); squares: auxiliary boundary knots for constructing B-splines.}
	\label{fig20}
\end{figure}

For example, the following is a cubic spline ($d = 4$) on $[a, b] = [1, 6]$, with interior knots at $s_1 = 2, s_2 = 3, s_3 = 4, s_4 = 5$:

\begin{equation*}
	f(x) = \begin{cases}
		1.09 + 0.610(x - 1) - 0.060(x - 1) ^ 2 - \rlap{$\frac{23}{75}$}{\phantom{\frac{000}{000}}}(x - 1) ^ 3 & x \in [1, 2),\\
		\rlap{$\frac{4}{3}$}{\phantom{0.00}} - 0.430(x - 2) - 0.980(x - 2) ^ 2 + \rlap{$\frac{59}{75}$}{\phantom{\frac{000}{000}}}(x - 2) ^ 3 & x \in [2, 3),\\
		0.71 - 0.030(x - 3) + 1.380(x - 3) ^ 2 - \frac{107}{150}(x - 3) ^ 3 & x \in [3, 4),\\
		\rlap{$\frac{101}{75}$}{\phantom{0.00}} + 0.590(x - 4) - 0.760(x - 4) ^ 2 + \rlap{$\frac{7}{24}$}{\phantom{\frac{000}{000}}}(x - 4) ^ 3 & x \in [4, 5),\\
		\rlap{$\frac{881}{600}$}{\phantom{0.00}} - 0.055(x - 5) + 0.115(x - 5) ^ 2 + \frac{37}{300}(x - 5) ^ 3 & x \in [5, 6).
	\end{cases}
\end{equation*}
To express $f(x)$ as a linear combination of B-splines, we need to arbitrarily place 3 auxiliary boundary knots on each side of $[1, 6]$. For example, with the following full knot sequence:
\begin{center}
	\begin{tabular}{|c|c|c|c|c|c|c|c|c|c|c|c|}
		& & & $a$ & $s_1$ & $s_2$ & $s_3$ & $s_4$ & $b$ & & &\\
		\hline
		$t_1$ & $t_2$ & $t_3$ &$t_4$ &$t_5$ &$t_6$ &$t_7$ &$t_8$ & $t_9$ & $t_{10}$ & $t_{11}$ & $t_{12}$\\
		\hline
		-2 & -1 & 0 & 1 & 2 & 3 & 4 & 5 & 6 & 7 & 8 & 9
	\end{tabular}
\end{center}
we can represent $f(x)$ using 8 uniform B-splines (see Figure \ref{fig20}) with coefficients 0.44, 1.11, 1.66, 0.25, 1.60, 1.43, 1.49 and 2.52. Placing auxiliary boundary knots elsewhere only results in different B-spline coefficients. For example, with clamped boundary knots $t_1 = t_2 = t_3 = t_4 = 1$ and $t_9 = t_{10} = t_{11} = t_{12} = 6$, B-spline coefficients are 1.09, $\frac{97}{75}$, 1.66, 0.25, 1.60, 1.43, 1.47 and $\frac{991}{600}$.

%\bigskip
\begin{center}
{\large\bf SUPPLEMENTARY MATERIAL}
\end{center}

\begin{description}

\item[R-code:] This document contains instructions to install \textbf{R} packages \textbf{gps} and \textbf{gps.mgcv}, and code to reproduce all demos, computation results, simulation outcomes and real data analysis in the paper. (.pdf file)

\end{description}

\bigskip
\begin{center}
	{\large\bf ACKNOWLEDGMENTS}
\end{center}

Zheyuan Li is grateful to Ahmed Elhakeem (Bristol Medical School, UK) and Adam Baxter-Jones (University of Saskatchewan, Canada), who shared their experience with modeling BMC data in our previous collaboration. BMC data are not open data. For this paper, a high-quality synthetic version is used and provided in \textbf{gps.mgcv}. Researchers who are interested in accessing the authentic data from PBMAS should contact Adam Baxter-Jones (baxter.jones@usask.ca).

\bigskip
\begin{center}
	{\large\bf FUNDING}
\end{center}

Zheyuan Li was supported by National Natural Science Foundation of China under Young Scientists Fund NSFC-12001166. Jiguo Cao was supported by Natural Sciences and Engineering Research Council of Canada under Discovery Grant 2018-06008.

\bigskip
\begin{center}
	{\large\bf ORCID}
\end{center}

Zheyuan Li, \url{https://orcid.org/0000-0002-7434-5947}

Jiguo Cao, \url{https://orcid.org/0000-0001-7417-6330}


\begin{thebibliography}{}

\bibitem[Andrinopoulou et~al., 2018]{WOS:000436403600033}
Andrinopoulou, E.-R., Eilers, P. H.~C., Takkenberg, J. J.~M., and Rizopoulos,
  D. (2018).
\newblock Improved dynamic predictions from joint models of longitudinal and
  survival data with time-varying effects using p-splines.
\newblock {\em Biometrics}, 74(2):685--693.

\bibitem[Arnold and Tibshirani, 2016]{genlasso-dual-path-implementation}
Arnold, T.~B. and Tibshirani, R.~J. (2016).
\newblock Efficient implementations of the generalized lasso dual path
  algorithm.
\newblock {\em Journal of Computational and Graphical Statistics}, 25(1):1--27.

\bibitem[Bailey, 1997]{PBMAS-study-original}
Bailey, D.~A. (1997).
\newblock The {S}askatchewan {P}ediatric {B}one {M}ineral {A}ccrual {S}tudy:
  Bone mineral acquisition during the growing years.
\newblock {\em International Journal of Sports Medicine}, 18:s191--s194.

\bibitem[Bates et~al., 2015]{lme4-paper}
Bates, D., M{\"a}chler, M., Bolker, B., and Walker, S. (2015).
\newblock Fitting linear mixed-effects models using {lme4}.
\newblock {\em Journal of Statistical Software}, 67(1):1--48.

\bibitem[Bates et~al., 2022]{lme4-package}
Bates, D., M{\"a}chler, M., Bolker, B., and Walker, S. (2022).
\newblock {\em {lme4}: Linear Mixed-Effects Models using 'Eigen' and S4}.
\newblock R package version 1.1-28.

\bibitem[Baxter-Jones et~al., 2011]{PBMAS-study}
Baxter-Jones, A.~D., Faulkner, R.~A., Forwood, M.~R., Mirwald, R.~L., and
  Bailey, D.~A. (2011).
\newblock Bone mineral accrual from 8 to 30 years of age: An estimation of peak
  bone mass.
\newblock {\em Journal of Bone and Mineral Research}, 26(8):1729--1739.

\bibitem[Bremhorst and Lambert, 2016]{WOS:000364259800022}
Bremhorst, V. and Lambert, P. (2016).
\newblock Flexible estimation in cure survival models using bayesian p-splines.
\newblock {\em Computational Statistics \& Data Analysis}, 93(SI):270--284.

\bibitem[Chen et~al., 2018]{WOS:000444443000004}
Chen, J., Ohlssen, D., and Zhou, Y. (2018).
\newblock Functional mixed effects model for the analysis of dose-titration
  studies.
\newblock {\em Statistics in Biopharmaceutical Research}, 10(3):176--184.

\bibitem[de~Boor, 2001]{deBoor-book}
de~Boor, C. (2001).
\newblock {\em A Practical Guide to Splines (Revised Edition)}, volume~27 of
  {\em Applied Mathematical Sciences}.
\newblock Springer New York.

\bibitem[DiMatteo et~al., 2001]{bayesian-free-knot-splines}
DiMatteo, I., Genovese, C.~R., and Kass, R.~E. (2001).
\newblock Bayesian curve-fitting with free-knot splines.
\newblock {\em Biometrika}, 88(4):1055--1071.

\bibitem[Eilers and Marx, 2021]{P-splines-book}
Eilers, P.~H. and Marx, B.~D. (2021).
\newblock {\em Practical smoothing: the joys of P-splines}.
\newblock Cambridge University Press.

\bibitem[Eilers and Marx, 1996]{P-splines}
Eilers, P. H.~C. and Marx, B.~D. (1996).
\newblock Flexible smoothing with {B}-splines and penalties.
\newblock {\em Statistical Science}, 11(2):89--102.

\bibitem[Eilers et~al., 2015]{P-splines-20-years}
Eilers, P. H.~C., Marx, B.~D., and Durbán, M. (2015).
\newblock Twenty years of {P}-splines.
\newblock {\em Statistics and Operations Research Transactions},
  39(2):149--186.

\bibitem[Elhakeem et~al., 2022]{BMC-trajectory-modeling}
Elhakeem, A., Hughes, R., Tilling, K., Cousminer, D., Jackowski, S., Cole, T.,
  Kwong, A., Li, Z., Grant, S., Baxter-Jones, A., Zemel, B., and Lawlor, D.
  (2022).
\newblock Using linear and natural cubic splines, sitar, and latent trajectory
  models to characterise nonlinear longitudinal growth trajectories in cohort
  studies.
\newblock {\em BMC Medical Research Methodology}, 22(68).

\bibitem[Franco-Villoria et~al., 2019]{WOS:000461592800005}
Franco-Villoria, M., Scott, M., and Hoey, T. (2019).
\newblock Spatiotemporal modeling of hydrological return levels: A quantile
  regression approach.
\newblock {\em Environmetrics}, 30(2, SI).

\bibitem[Gervini, 2006]{free-knot-smoothing-functional-data}
Gervini, D. (2006).
\newblock Free-knot spline smoothing for functional data.
\newblock {\em Journal of the Royal Statistical Society: Series B (Statistical
  Methodology)}, 68(4):671--687.

\bibitem[Gijbels et~al., 2018]{WOS:000434068100003}
Gijbels, I., Ibrahim, M.~A., and Verhasselt, A. (2018).
\newblock Testing the heteroscedastic error structure in quantile varying
  coefficient models.
\newblock {\em Canadian Journal of Statistics}, 46(2):246--264.

\bibitem[Goicoa et~al., 2019]{WOS:000456529100005}
Goicoa, T., Adin, A., Etxeberria, J., Militino, A.~F., and Ugarte, M.~D.
  (2019).
\newblock Flexible bayesian p-splines for smoothing age-specific
  spatio-temporal mortality patterns.
\newblock {\em Statistical Methods in Medical Research}, 28(2):384--403.

\bibitem[Greco et~al., 2018]{WOS:000448217800001}
Greco, F., Ventrucci, M., and Castelli, E. (2018).
\newblock P-spline smoothing for spatial data collected worldwide.
\newblock {\em Spatial Statistics}, 27:1--17.

\bibitem[Hendrickx et~al., 2018]{WOS:000450660500010}
Hendrickx, K., Janssen, P., and Verhasselt, A. (2018).
\newblock Penalized spline estimation in varying coefficient models with
  censored data.
\newblock {\em TEST}, 27(4):871--895.

\bibitem[Jupp, 1978]{approximation-to-data-by-splines-with-free-knots}
Jupp, D. L.~B. (1978).
\newblock Approximation to data by splines with free knots.
\newblock {\em SIAM journal on numerical analysis}, 15(2):328--343.

\bibitem[Koehler et~al., 2017]{WOS:000418746100004}
Koehler, M., Umlauf, N., Beyerlein, A., Winkler, C., Ziegler, A.-G., and
  Greven, S. (2017).
\newblock Flexible bayesian additive joint models with an application to type 1
  diabetes research.
\newblock {\em Biometrical Journal}, 59(6, SI):1144--1165.

\bibitem[Li and Cao, 2022a]{gps-package}
Li, Z. and Cao, J. (2022a).
\newblock {\em {gps}: General P-Splines}.
\newblock R package version 1.1.

\bibitem[Li and Cao, 2022b]{gps.mgcv-package}
Li, Z. and Cao, J. (2022b).
\newblock {\em {gps.mgcv}: General P-Splines for Package 'mgcv'}.
\newblock R package version 1.0.

\bibitem[Lindstrom, 1999]{penalized-free-knot-splines}
Lindstrom, M.~J. (1999).
\newblock Penalized estimation of free-knot splines.
\newblock {\em Journal of computational and graphical statistics},
  8(2):333--352.

\bibitem[Mammen and van~de Geer, 1997]{locally-adaptive-regression-splines}
Mammen, E. and van~de Geer, S. (1997).
\newblock Locally adaptive regression splines.
\newblock {\em The Annals of statistics}, 25(1):387--413.

\bibitem[Maturana-Russel and Meyer,
  2021]{spectral-density-estimation-using-P-splines-with-quantile-knots}
Maturana-Russel, P. and Meyer, R. (2021).
\newblock Bayesian spectral density estimation using {P}-splines with
  quantile-based knot placement.
\newblock {\em Computational statistics}, 36(3):2055--2077.

\bibitem[Minguez et~al., 2020]{WOS:000489511900001}
Minguez, R., Basile, R., and Durban, M. (2020).
\newblock An alternative semiparametric model for spatial panel data.
\newblock {\em Statistical Methods and Applications}, 29(4):669--708.

\bibitem[Miyata and Shen, 2003]{adaptive-free-knot-splines}
Miyata, S. and Shen, X. (2003).
\newblock Adaptive free-knot splines.
\newblock {\em Journal of Computational and Graphical Statistics},
  12(1):197--213.

\bibitem[Muggeo et~al., 2021]{WOS:000549910800001}
Muggeo, V. M.~R., Torretta, F., Eilers, P. H.~C., Sciandra, M., and Attanasio,
  M. (2021).
\newblock Multiple smoothing parameters selection in additive regression
  quantiles.
\newblock {\em Statistical Modelling}, 21(5):428--448.

\bibitem[Orbe and Virto, 2021]{WOS:000641387600001}
Orbe, J. and Virto, J. (2021).
\newblock Selecting the smoothing parameter and knots for an extension of
  penalized splines to censored data.
\newblock {\em Journal of Statistical Computation and Simulation},
  91(14):2953--2985.

\bibitem[O'Sullivan, 1986]{O-splines-1}
O'Sullivan, F. (1986).
\newblock A statistical perspective on ill-posed inverse problems.
\newblock {\em Statistical Science}, 1(4):502--518.

\bibitem[Pedersen et~al., 2019]{hierarchical-GAM}
Pedersen, E.~J., Miller, D.~L., Simpson, G.~L., and Ross, N. (2019).
\newblock Hierarchical generalized additive models in ecology: an introduction
  with mgcv.
\newblock {\em PeerJ}, 7:e6876--e6876.

\bibitem[Perperoglou et~al., 2019]{review-of-R-packages-on-splines-2019}
Perperoglou, A., Sauerbrei, W., Abrahamowicz, M., and Schmid, M. (2019).
\newblock A review of spline function procedures in r.
\newblock {\em BMC medical research methodology}, 19(1):46--46.

\bibitem[Pinheiro et~al., 2022]{nlme-package}
Pinheiro, J., Bates, D., DebRoy, S., Sarkar, D., and {R Core Team} (2022).
\newblock {\em {nlme}: Linear and Nonlinear Mixed Effects Models}.
\newblock R package version 3.1-157.

\bibitem[Pinheiro and Bates, 2000]{linear-mixed-models-book}
Pinheiro, J.~C. and Bates, D.~M. (2000).
\newblock {\em Mixed-Effects Models in S and S-PLUS}.
\newblock Springer.

\bibitem[Ramdas and Tibshirani, 2016]{trendfilter-ADMM}
Ramdas, A. and Tibshirani, R.~J. (2016).
\newblock Fast and flexible admm algorithms for trend filtering.
\newblock {\em Journal of computational and graphical statistics},
  25(3):839--858.

\bibitem[Rodriguez-Alvarez et~al., 2018]{WOS:000426321800004}
Rodriguez-Alvarez, M.~X., Boer, M.~P., van Eeuwijk, F.~A., and Eilers, P. H.~C.
  (2018).
\newblock Correcting for spatial heterogeneity in plant breeding experiments
  with p-splines.
\newblock {\em Spatial Statistics}, 23:52--71.

\bibitem[Ruppert et~al., 2003]{SemiPar-book-2003}
Ruppert, D., Wand, M.~P., and Carroll, R.~J. (2003).
\newblock {\em Semiparametric Regression}.
\newblock Cambridge Series in Statistical and Probabilistic Mathematics.
  Cambridge University Press.

\bibitem[Ruppert et~al., 2009]{SemiPar-article-2009}
Ruppert, D., Wand, M.~P., and Carroll, R.~J. (2009).
\newblock Semiparametric regression during 2003-2007.
\newblock {\em Electronic journal of statistics}, 3:1193--1256.

\bibitem[Segal et~al., 2018]{L1-P-splines}
Segal, B.~D., Elliott, M.~R., Braun, T., and Jiang, H. (2018).
\newblock P-splines with an {L}1 penalty for repeated measures.
\newblock {\em Electronic journal of statistics}, 12(2):3554--3600.

\bibitem[Silverman, 1985]{Silverman-some-aspects-of-smoothing-splines}
Silverman, B.~W. (1985).
\newblock Some aspects of the spline smoothing approach to non-parametric
  regression curve fitting.
\newblock {\em Journal of the Royal Statistical Society: Series B (Statistical
  Methodology)}, 47(1):1--52.

\bibitem[Spiegel et~al., 2019]{WOS:000457464800009}
Spiegel, E., Kneib, T., and Otto-Sobotka, F. (2019).
\newblock Generalized additive models with flexible response functions.
\newblock {\em Statistics and Computing}, 29(1):123--138.

\bibitem[Spiegel et~al., 2020]{WOS:000546038500003}
Spiegel, E., Kneib, T., and Otto-Sobotka, F. (2020).
\newblock Spatio-temporal expectile regression models.
\newblock {\em Statistical Modelling}, 20(4):386--409.

\bibitem[Spiriti et~al., 2013]{freeknotsplines-archived-by-CRAN}
Spiriti, S., Eubank, R., Smith, P.~W., and Young, D. (2013).
\newblock Knot selection for least-squares and penalized splines.
\newblock {\em Journal of statistical computation and simulation},
  83(6):1020--1036.

\bibitem[Tibshirani, 2014]{polynomial-trendfilter}
Tibshirani, R.~J. (2014).
\newblock Adaptive piecewise polynomial estimation via trend filtering.
\newblock {\em The Annals of statistics}, 42(1):285--323.

\bibitem[Tibshirani and Taylor, 2011]{genlasso-dual-path-theory}
Tibshirani, R.~J. and Taylor, J. (2011).
\newblock The solution path of the generalized lasso.
\newblock {\em The Annals of statistics}, 39(3):1335--1371.

\bibitem[Wahba, 1990]{Wahba-spline-models-book}
Wahba, G. (1990).
\newblock {\em Spline models for observational data}.
\newblock Society for Industrial and Applied Mathematics.

\bibitem[Wand, 2018]{SemiPar-package}
Wand, M. (2018).
\newblock {\em {SemiPar}: Semiparametic Regression}.
\newblock R package version 1.0-4.2.

\bibitem[Wand and Ormerod, 2008]{SemiPar-with-O-splines}
Wand, M.~P. and Ormerod, J.~T. (2008).
\newblock On semiparametric regression with {O}'{S}ullivan penalized splines.
\newblock {\em Australian \& New Zealand journal of statistics},
  50(2):179--198.

\bibitem[Wang et~al., 2018]{WOS:000440611000006}
Wang, X., Roy, V., and Zhu, Z. (2018).
\newblock A new algorithm to estimate monotone nonparametric link functions and
  a comparison with parametric approach.
\newblock {\em Statistics and Computing}, 28(5):1083--1094.

\bibitem[Wood, 2022]{mgcv-package}
Wood, S. (2022).
\newblock {\em {mgcv}: Mixed GAM Computation Vehicle with Automatic Smoothness
  Estimation}.
\newblock R package version 1.8-40.

\bibitem[Wood, 2017a]{Wood-GAMs-book}
Wood, S.~N. (2017a).
\newblock {\em Generalized Additive Models: An Introduction with R}.
\newblock Chapman and Hall, 2nd edition.

\bibitem[Wood, 2017b]{mgcv-B-splines}
Wood, S.~N. (2017b).
\newblock P-splines with derivative based penalties and tensor product
  smoothing of unevenly distributed data.
\newblock {\em Statistics and Computing}, 27(4):985--989.

\bibitem[Yao and Lee, 2008]{YaoFang-knot-placement}
Yao, F. and Lee, T. C.~M. (2008).
\newblock On knot placement for penalized spline regression.
\newblock {\em Journal of the Korean Statistical Society}, 37(3):259--267.

\bibitem[Yu et~al., 2017]{WOS:000395004300017}
Yu, Y., Wu, C., and Zhang, Y. (2017).
\newblock Penalised spline estimation for generalised partially linear
  single-index models.
\newblock {\em Statistics and Computing}, 27(2):571--582.

\bibitem[Zuur et~al., 2014]{highland-GAMMs-book}
Zuur, A.~F., Saveliev, A.~A., and Ieno, E.~N. (2014).
\newblock {\em A Beginner's Guide to Generalized Additive Mixed Models with R}.
\newblock Highland Statistics.

\end{thebibliography}
\end{document}